\documentclass[final,5p,times,twocolumn]{elsarticle}

\usepackage{graphicx}
\usepackage{amssymb}
\usepackage{amsthm}
\usepackage{amsmath}
\usepackage{float}
\usepackage{dblfloatfix}
\usepackage{lineno}
\usepackage{subfigure}

\usepackage{upgreek}

\biboptions{sort&compress}

\usepackage{booktabs}

\usepackage{mathptmx}
\usepackage{lscape}

\usepackage{array}
\usepackage{droidsans}
\usepackage{charter}
\usepackage[usenames,dvipsnames]{xcolor}
\usepackage{setspace}
\usepackage[colorlinks]{hyperref}
\usepackage{chemformula}
\usepackage[detect-none]{siunitx}
\usepackage{booktabs}
\usepackage{caption}
\usepackage{epstopdf}

\journal{Journal of molecular liquids}

\begin{document}
 
\begin{frontmatter}

\title{Stockmayer supracolloidal magnetic polymers under the influence of an applied magnetic field and a shear flow}

\author[vienna]{Ivan S. Novikau}
\author[russia]{Vladimir V. Zverev}
\author[russia]{Ekaterina V. Novak}
\ead{ekaterina.novak@urfu.ru}
\author[vienna,MMM]{Sofia S. Kantorovich}

\address[vienna]{Faculty of Physics, University of Vienna, Kolingasse 14-16, 1090 Vienna, Austria}
\address[russia]{Ural Federal University, Lenin Av. 51, Ekaterinburg 620000, Russian Federation}
\address[MMM]{Research Platform ``Mathematics-Magnetism-Materials'', University of Vienna, Oskar-Morgenstern-Platz 1, 1090 Vienna, Austria }

\begin{abstract}

The idea of creating magnetically controllable colloids whose rheological properties can be finely tuned on the nano- or micro-scale has caused a lot of experimental and theoretical effort. The latter resulted in systems whose building blocks are ranging between single magnetic nanoparticles to complexes of such nanoparticles bound together by various mechanisms. The binding can be either chemical or physical, reversible or not. One way to create a system that is physically bound is to let the precrosslinked supracolloidal magnetic polymers (SMPs) to cluster due to both magnetic and Van-der-Waals-type forces. The topology of the SMPs in this case can be used to tune both magnetic and rheological properties of the resulting clusters as we show in this work. We employ Molecular Dynamics computer simulations coupled with explicit solvent modelled by Lattice-Boltzmann method in order to model the behaviour of the clusters formed by chains, rings, X- and Y-shaped SMPs in a shear flow with externally applied magnetic field. We find that the shear stabilises the shape of the clusters not letting them extend in the direction of the field and disintegrate. The clusters that show the highest response to an applied field and higher shape stability are those made of Y- and X-like SMPs.
\end{abstract}

\end{frontmatter}

\section{Introduction}\label{sec:intro}

Even the basic magnetic soft matter system -- magnetic fluid, {\it i.e.} a suspension of magnetic nanoparticles in magneto-passive carriers -- that was synthesised more than 55 years ago \cite{resler1964magnetocaloric} still possesses properties that are not well understood. This is mainly related to the dynamics and rheology, as they are heavily affected by nanoparticle polydispersity, concentration, interactions and applied magnetic fields \cite{maldonado16a,remmer17a,ivanov19a,lebedev19a,angbo20a,camp21a,kuznetsov22a}. Nonetheless, it is the magnetic off-equilibrium dynamic properties that make magnetic soft matter so appealing for medical and technological applications \cite{2004-menager-pol,2015-backes-jpcb,rosenberg20a,mandal19a,rosenberg22a,2012-ren,sung20a,cao20a,gao20a,biglione20a,hilger12iron,durr2013magnetic,perigo2015fundamentals,ABENOJAR2016440}. In order to optimise the aforementioned applications, the researchers came out with various modifications of magnetic component: increasing the size to microns, this way creating magnetorheological suspensions \cite{cutillas98a,furst00a,melle03a,bossis03a}, or changing particle shape and internal structure \cite{Meyer04,Tan05,Lisjak18,Castanheira20,Saccanna11,Rossi11,2014-tierno,donaldson2017cube,Mertelj22}, creating multicore magnetic particles \cite{eberbeck2013multicore,ludwig2014magnetic,trisnanto2021effective} or emulsions with droplets of dense magnetic phase \cite{cunha18a,zhang19a,abicalil21a,radcliffe21a,shunichi22a}, employing magneto-elastic coupling in soft colloids \cite{2019-witte-sm,mandal19a,novikau2020influence}  or supramolecular polymeric structures \cite{1998-furst,2003-goubault,2008-benkoski,2009-zhou,2011-benkoski,Dreyfus_2005,bereczk2017biotemplated,kralj2015magnetic,nanoletter_2021, mostarac2020characterisation}. The common idea of the last three systems is to use chemical or physical bonds in order to create small nanoparticle clusters that, in contrast to their counterparts formed by means of self-assembly, are stable even if external stimuli such as magnetic field, heat or hydrodynamic flow are applied.

One of the possibilities to create complex colloids made of multiple nanoparticles is build on the van-der-Waals-type interactions. It is known that, in case of additional central attraction between the magnetic dipolar particles (Stockmayer fluids), instead of self-assembling into loose linear or branched clusters as their counterparts without central attraction do \cite{jordan1999magnetic,2000-camp-prl,1999-levin,2005-holm,kantorovich2013nonmonotonic,rovigatti2013branching,2015-kantorovich-pccp1}, such systems undergo gas-liquid phase transition with particles forming droplet-like compact aggregates \cite{1989-smit,1993-vanleeuwen,1992-panagiotopoulos,1995-stevens,1981-adams}. Recently, we showed the advantage of using Stockmayer supracolloidal magnetic polymer-like structures (SSMPs) to fine-tune the magnetic response of formed compact aggregates: it turned out that depending on the topology of the individual SSMPs, the initial susceptibility of the droplets they form can be tuned. Moreover, the critical applied field needed to deform/disintegrate the cluster  was found to be SSMP-topology-dependent as well \cite{zverev2021,novak2019}. All those findings, however, correspond to equilibrium states, and were obtained fully ignoring the effects of hydrodynamics, thus, not shedding light on the rheology of the clusters. In order to use complex colloids in micro(nano)fluidics for drug delivery, for instance, three key parameters are to be optimised: cargo capacity, transport properties and the ability to control those two externally.

\begin{figure}[h!]
    \centering
    \includegraphics[width=\linewidth]{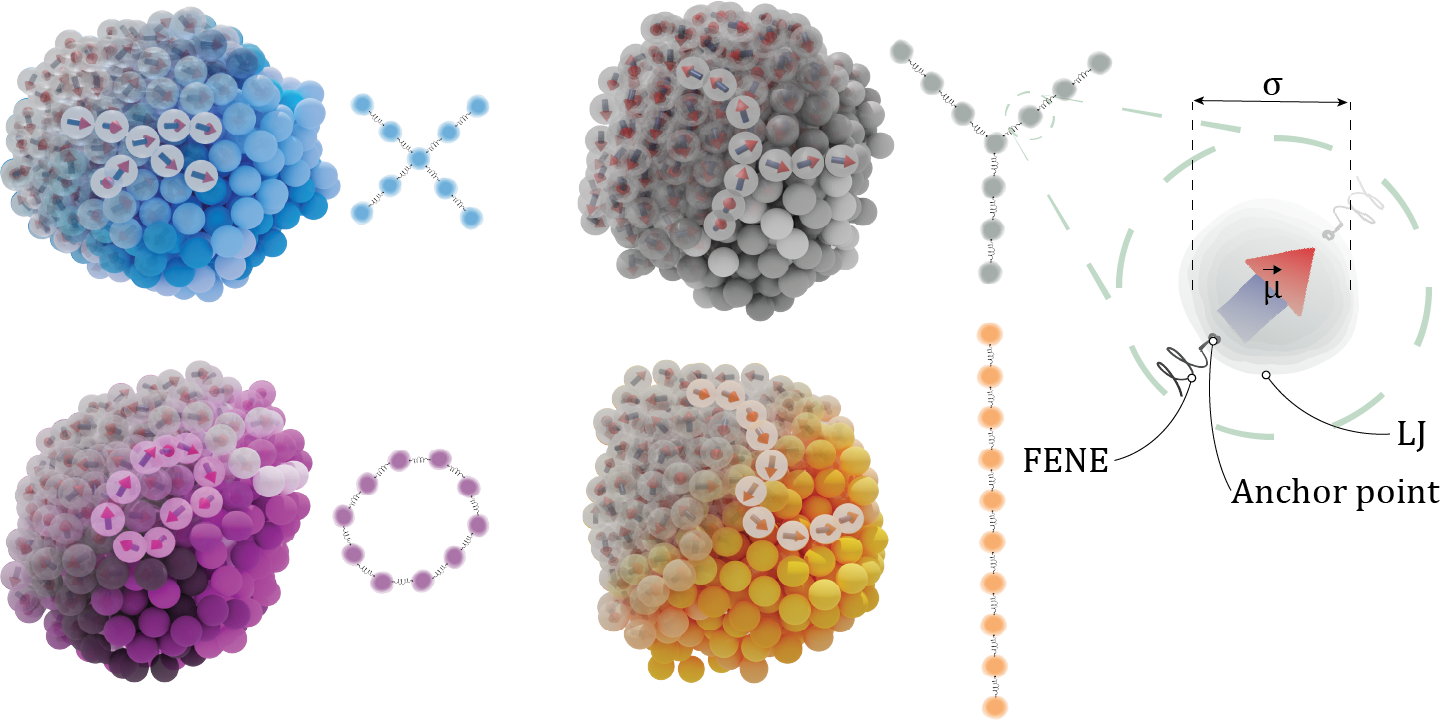}
    \caption{Equilibrium clusters formed by X-, Y-, chain- and ring-like SSMPs, arrows indicate point dipoles; In the inset one can find a core-shell magnetic nanoparticle (MNP). The bonding between crosslinked monomers within every polymer is modelled as a FENE potential. FENE-like spring attachment points are placed at the projection points of the head and the tail of the central dipole moment.}
    \label{fig:scheme}
\end{figure}

In this study we employ molecular dynamics computer simulations at fixed temperature, $T$, combined with Lattice-Boltzmann approach \cite{2009-duenweg-lb} to investigate the magneto-rheological in-field behaviour of the clusters formed by Y-, X-, chain- and ring-like SSMPs as shown in Fig. \ref{fig:scheme}. Those clusters are obtained in the thermodynamic equilibrium and as such are stable if no external forces are applied. Each magnetic particle has a point dipole, $\vec{\mu}$, whose orientation, shown with arrows, is driven by dipolar forces and frustration caused by the central attraction.

\begin{figure}[h!]
 \centering
    \includegraphics[width=0.75\linewidth]{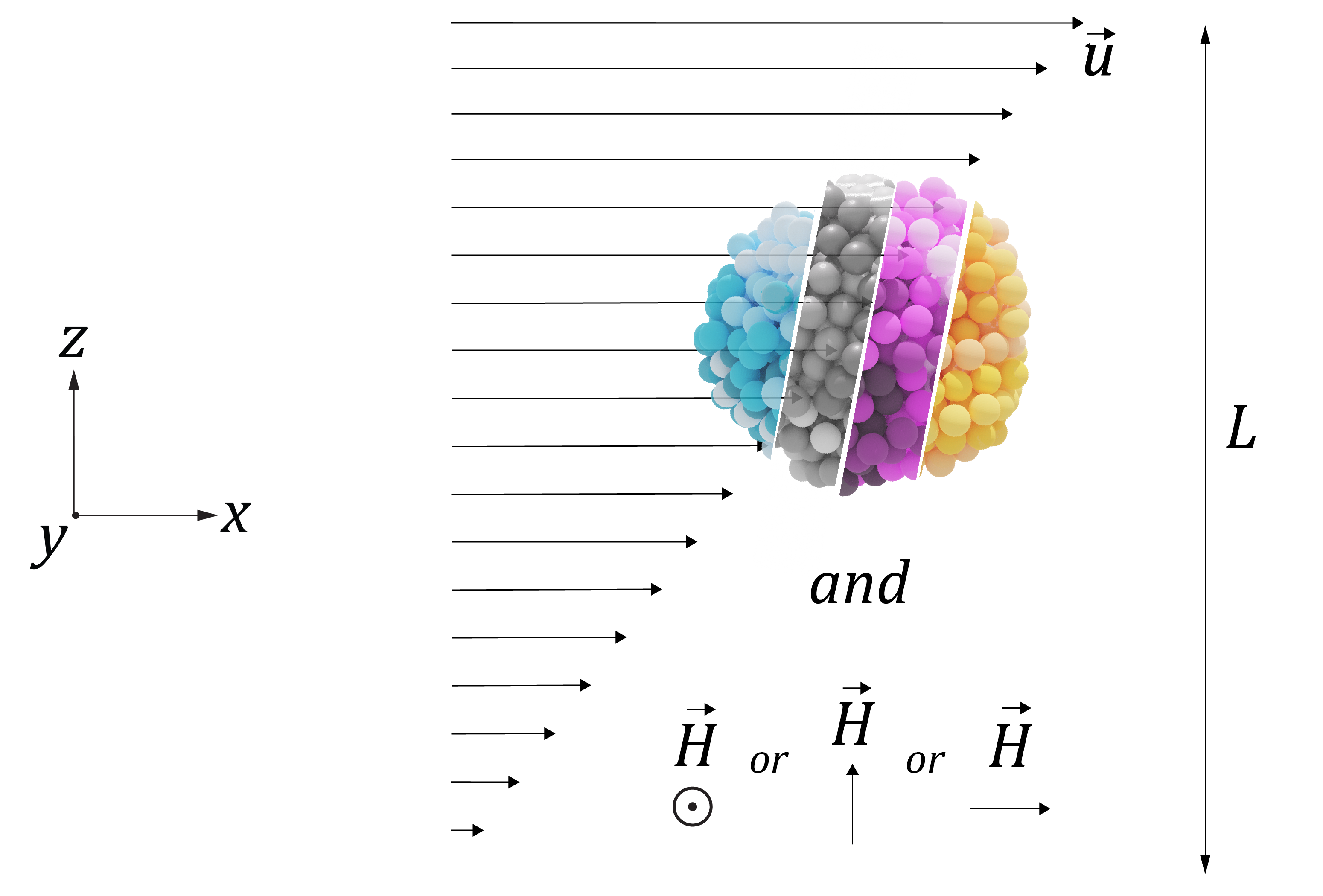}
    \caption{Geometry of the system: the upper wall of a slit with the height
    $L$ moves with the velocity $u$. Equilibrated cluster of a given topology (X, Y, O or I) is placed inside and let evolve. Magnetic field is applied along \textit{x-}, \textit{y-}, or \textit{z-}axis.}
    \label{fig:sys-geom}
\end{figure}

We put each of this clusters in a slit of height $L$  with a shear flow, and apply an external magnetic field along one of the three axis as illustrated in Fig. \ref{fig:sys-geom}. We first investigate how the flow and field affect the shape of the clusters and if a particular combination of the field-flow orientations facilitates the disentanglement and further disintegration of the clusters (see, SupVideo1) more than the others. This way, we characterise cargo abilities of the clusters. Next, we carefully analyse the dynamics of the clusters in various flow-field combinations and elucidate the influence of the SSMP topology on the cluster motion in order to verify cluster transport properties. Finally, we analyse the magnetisation of the clusters both as a function of the field-flow combination and time, aiming at establishing a protocol to control these objects. 

\section{Simulation Approach}\label{ref:sim_app}

In this section we provide all the details of the simulation methods employed to analyse magnetorheological behaviour of the SSMP clusters.

\subsection{Simulation Protocol}\label{subs:sim_protocol}

Our main tool is the molecular dynamics \cite{2002-frenkel} on a coarse-grained level: all degrees of freedom up to molecular ones are integrated out and each colloidal bead of an SSMP is represented by a sphere. 
In order to investigate the effects of the shear flow on such systems, 
one needs to have an approach for modelling the solvent. 
Among various possible techniques \cite{lusebrink16a,cerda19a,2019-rovigatti-sm-rev,weiss2019spatial,formanek21a}, 
we opt for the Lattice-Boltzmann (LB) scheme \cite{pagonabarraga2004lattice,2008-dunweg}. 
This method simulates the flow of a Newtonian fluid by solving the discrete Boltzmann equation 
on a lattice, which corresponds to the Navier-Stokes equations in the limit of small Mach numbers. The background fluid itself is represented by virtual ``effective'' 
particles that are distributed in physical space on the nodes of a regular lattice, 
taking the form of discrete velocity and density distribution functions.

In previous
work \cite{NOVIKAU2022118056}, we already successfully checked the correctness of our LB code on a benchmark system, by comparing the results to the 
solution of the steady–state Navier–Stokes equation, utilising a finite volume solver over a non-uniform 
body-fitted structured mesh.

For the coupling of SSMP beads to the fluid, 
we use the  scheme of Ahlrichs and D\"unweg \cite{2008-dunweg}. In this approach the system is thermalised by the fluid. 
It is worth mentioning, however, that each bead is a point particle, meaning 
that its rotation is invisible for the fluid. While being a reasonable approximation for 
models of polymers \cite{kreissl2021frequency}, this limitation is crucial for adequate simulation of magnetic nanoparticles (MNPs).

We overcome it by initializing
Brownian dynamics for thermalisation of dipoles rotation. Additionally, we ensured that the Brownian rotation frequency of our simulated MNPs matched that of real ones with similar hydrodynamic radii and viscosity conditions. This approach neglects the lubrication effect in the hydrodynamic interactions between MNPs, which is acceptable when the hydrodynamic forces are not excessively large and a polymeric shell surrounds the MNPs, providing sufficient separation between their surfaces. Furthermore, the significance of torque exchange between MNPs through many-body hydrodynamic interactions is relatively modest \cite{kim2009hydrodynamic}. Magnetic interactions between MNPs are calculated by direct summation. 
We use code developed under ESPResSo simulation package (ver. 4.1.4) \cite{Weik2019ESPResSoSystems}. 

As shown in Fig. \ref{fig:sys-geom}, we simulate clusters in a slit formed between two walls separated by distance 
$L$ along $z$-axis. A linear shear fluid velocity profile, with corresponding shear rate $\dot{\gamma}$, 
was created by the moving upper wall along $x$-axis. 
Shear rates are set to $3$, $6$ or $9\times10^{5} s^{-1}$ . The range of magnetic field magnitudes lies between $67.5$ and $268.5$ mT, and
it is directed along \textit{x-}, \textit{y-} or \textit{z-}axis.  
Initially, every cluster is placed in the slit center in its flow-free equilibrium configuration and let to evolve in time. 

Cube-shaped simulation box with a side length of $L = 100 \sigma$ was used. The LB lattice constant is set 
to $a_{grid} = 1 \sigma$.
The system propagates with a fixed time step $\delta t=0.01$, updating LB fluid field at every second MD step. 
When analysing the data, the first  $2\times 10^{6}$ simulation steps 
are not considered. This period of time approximately corresponds to the first 1-2 rotations of the cluster under the influence of the shear. 
The total duration of an average simulation is about $7.5\times10^{6}$ steps. All measurements are averaged over four different 
cluster configurations with distinct SMPs distributions, 
in order to avoid dependence on the individual self-assembled version of a given topology. 

\subsection{SSMPs and interactions}

We use the same way of building SSMPs as in our previous work \cite{zverev2021}: first, single particles are crosslinked into SSMPs of linear, Y-, X- and ring-like shapes.  Zooming in any selected bead, with an effective diameter, $\sigma$,  as shown in the inset in Fig. \ref{fig:scheme}, one can find a core-shell magnetic nanoparticle (MNP) with mostly $2$ springs attached to it, except for free ends with a single spring, and central beads in X- and Y- SSMPs, having $4$ and $3$ anchor points respectively. Each bead mimics cobalt ferrite ($\mathrm{CoFe_2O_4}$) MNP of $\sim 20$ nm in diameter with $2$ nm in thickness polymeric shell, which is required for steric stabilisation. 

Their steric interactions are a combination of the central attraction and short range repulsion, modelled via  Lennard-Jones potential. By taking its energy scale as unity, the latter can be written as follows:
\begin{equation}
U_{\mathrm{{LJ}}}(r)=4 \left ( r^{-12}-r^{-6} \right ), 
\label{eq:LJ}
\end{equation}
\noindent where $r$ is the distance between centres of two interacting beads measured in units of $\sigma$. Here, we assume the ratio between the well of the LJ potential and the thermal energy to be unity.   To use potential from Eq. \eqref{eq:LJ} is a common practice to approximate van der Waals forces in coarse-grained molecular dynamics simulations \cite{2002-frenkel}. 

In the center of every bead there is  a point dipole with fixed direction in the body frame of reference and a constant length. 
The size-range considered in this paper justifies this approach, as cobalt ferrite nanoparticles 
of 20.1 nm in diameter will relax following the Brownian mechanism, 
rather than N{\'e}el one \cite{ota19a}. The interaction between two magnetic particles can be, 
thus, described by the dipole-dipole potential:

\begin{equation}
U_{d d}\left(\vec{r}_{i j}\right)=\frac{\left(\vec{\mu}_{i} \cdot \vec{\mu}_{j}\right)}{r^{3}}-\frac{3\left(\vec{\mu}_{i} \cdot \vec{r}_{i j}\right)\left(\vec{\mu}_{j} \cdot \vec{r}_{i j}\right)}{r^{5}},
\label{eq:dipdip}
\end{equation}

\noindent where $\vec \mu_i$, $\vec \mu_j$ are the respective dipole moments of particles 
that interact and $\vec r_{ij}$ is the vector connecting their centers. 
The standard way to characterize the interaction between two MNPs is to use the dipolar coupling parameter 
$\lambda$, which is the ratio of the maximum dipole-dipole interaction energy at contact to the thermal energy, 
what can be defined as $\lambda=\mu^2$. In this study we have selected MNPs in such a way that $\lambda=5$.

The field, $\vec H$, couples to each monomer dipole moment, $\mu$, via Zeeman potential:
\begin{equation}
 U_{\mathrm{Z}}(\vec \mu, \vec H)=-\vec \mu \cdot \vec H.
\end{equation}
Thus, an external field tends to coalign each dipole with its direction, whereas central attraction tries to maximise the number of nearest neighbours.

The combination of Lennard-Jones and dipolar interactions are usually addressed as Stockmayer interactions.

The polymer-like structure that we call Stockmayer supracolloidal magnetic polymer (SSMP) is build out of ten beads in case of linear, ring- and Y-like topologies, while X topology contain only 9 beads. Each topology is highlighted in Fig. \ref{fig:sys-geom}  inside a cluster and is sketched  next to it.
Beads inside one SSMP are interconnected by means of finitely 
extensible nonlinear elastic (FENE) springs tied to anchor points on the surface of beads, creating the SMPs backbones. 
If a bead has two FENE springs connected to it, then anchor points are diametrically opposed. In case of four anchor points (central bead of X topology), they are evenly distributed on beads equatorial circle, and in case of three anchor points (central bead of Y topology), central 
angles between adjacent anchor points are 135°, 90° and 135°. 
FENE potential has the form:
\begin{equation}
U_{F E N E}(r)=-\frac{1}{2} \epsilon_{f} r_f^2 \ln \left[1-\left(\frac{r}{r_{f}}\right)^{2}\right],
\label{eq:fene}
\end{equation}
\noindent with the dimensionless bond rigidity $\epsilon_{f}=100$ and the maximum bond extension $r_{f}=1.02$.

\subsection{Dimensionless Units}\label{subs:units}

\begin{figure}[h!]
    \centering
    \includegraphics[width=0.9\linewidth]{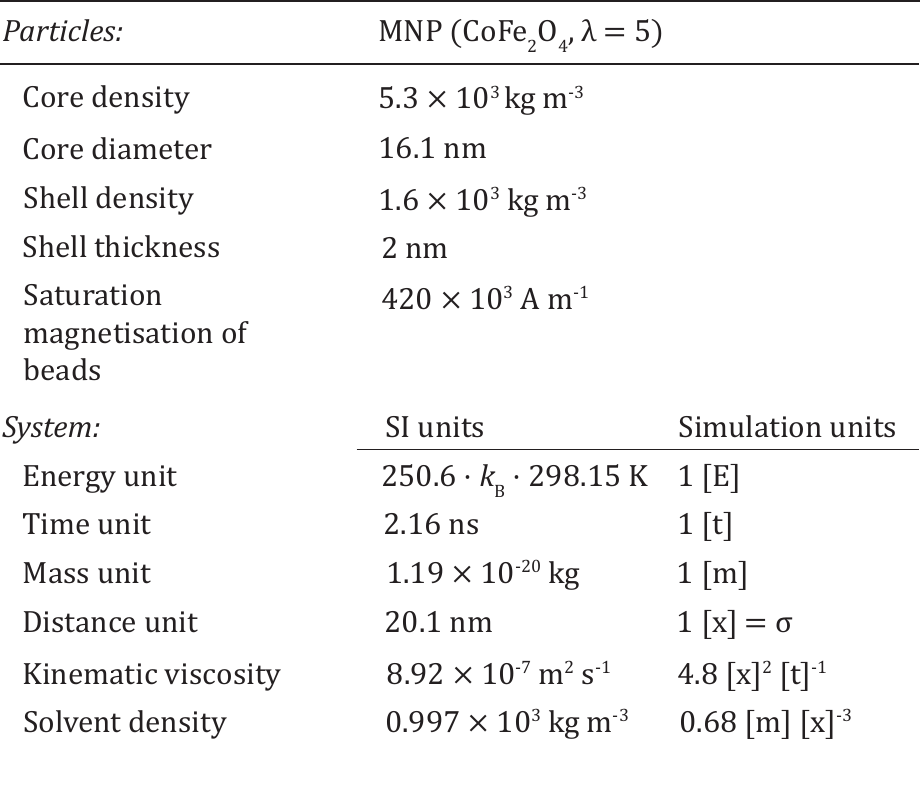}
    \captionof{table}{MNP's and system parameters in SI and their corresponding values in simulation units. 
}
    \label{fig:table}
\end{figure}

For computational convenience, it makes sense to use simulation unit (SU) system. A bead has dimensionless unit diameter $\sigma = 1$, what defines the length scale in the simulation. The Table \ref{fig:table} shows values of simulation parameters in SI an SU units.

\section{Results and Discussions}\label{ref:rnd}

\subsection{Cluster Deformation and Motion}\label{subs:def-av}

We start our analysis by measuring the radius of  gyration, $R_g$, of the clusters as a function of an applied magnetic field:
\begin{figure*}[h!]
    \centering
    \includegraphics[trim={0 7.0cm 0 0},clip=true, width=0.8\linewidth]{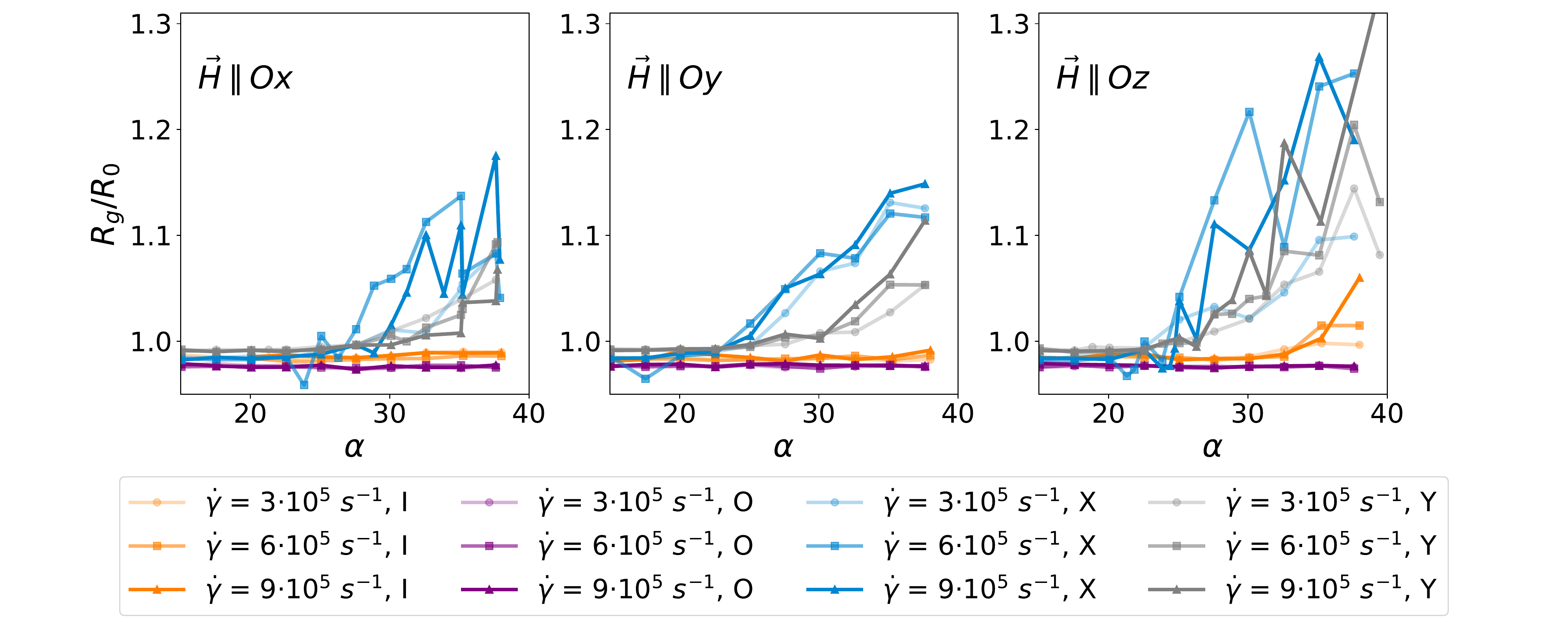}
    \caption{Relative radius of gyration, $R_g/R_0$, $R_0=R_g(0)$, as a function of the Langevin parameter $\alpha$. $R_0$ is calculated in the absence of an applied field and zero shear rate. Three panels correspond to different orientations of the applied field as shown in their bodies from the left to the right:  the field is parallel to the flow velocity, $\vec{H}\parallel Ox$; the field is perpendicular both to the flow velocity gradient and to the flow velocity, $\vec{H} \parallel Oy$; the field along the flow velocity gradient, $\vec{H}\parallel Oz$. For each SSMP topology, differentiated by colours -- blue for X-like, grey for Y-like, yellow for linear and purple for rings -- the curves for three different shear rates, $\dot{\gamma}$ are plotted as detailed in the legend below the plots.} \label{fig:RgH}
\end{figure*}

\begin{equation}
 R_g (\alpha) = \left [ \frac{1}{2N_p} \sum_{i,j=1}^{N_p} (\vec r_i - \vec r_j)^2\right ]^{1/2} ,
 \label{eq:Rg}
\end{equation}

\noindent where dimensionless parameter $\alpha = |\vec{\mu}| |\vec{H}|/(k_BT)$ shows the ratio between Zeeman and thermal energy, $k_B$ -- Boltzmann constant. The summation is carried over all $N_p$ monomers of a cluster, located at positions $\vec{r}_i$. As long as the system is off-equilibrium, the time average might only have a meaning in case of a steady state. So, we let the system to establish a steady flow profile, for clusters made of linear and ring-like  SSMPs we discard the first 2$\times10^4$ integration steps, while for clusters of Y- and X-like SSMPs the offset was  5$\times10^4$ integration steps, and then average $ R_g (\alpha)$ over time. The results are plotted in Fig. \ref{fig:RgH}.

\begin{figure*}[h!]
    \centering
         \subfigure[]{\label{fig:qh} 
         \includegraphics[width=0.7\linewidth]{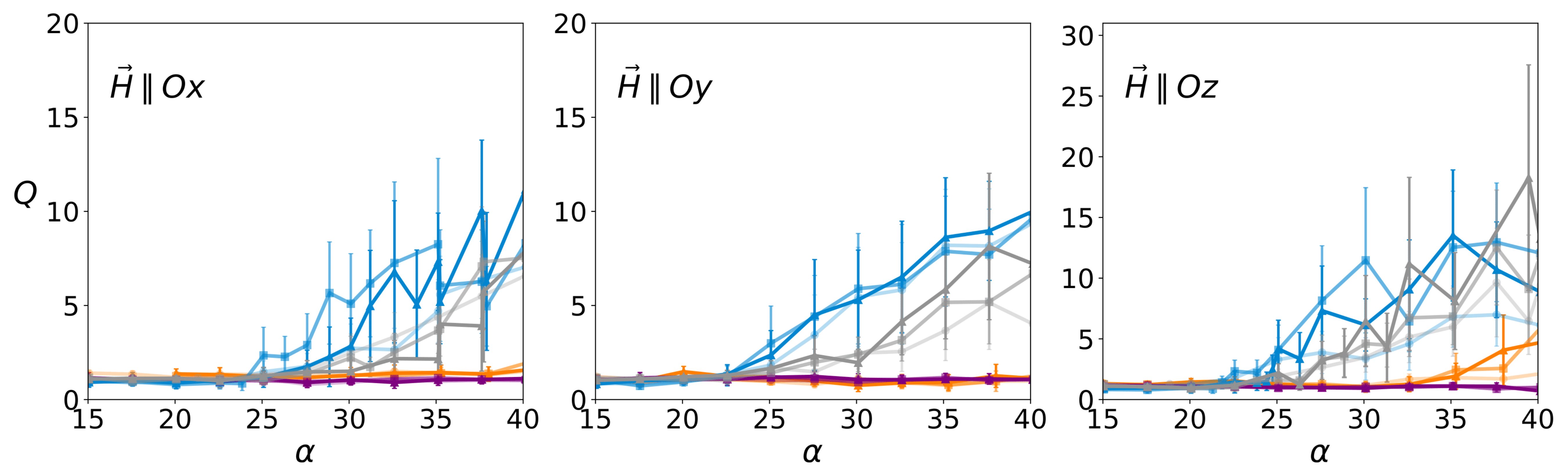}}
    \hfill
         \subfigure[]{\label{fig:all-field-orient}
         \includegraphics[width=0.7\linewidth]{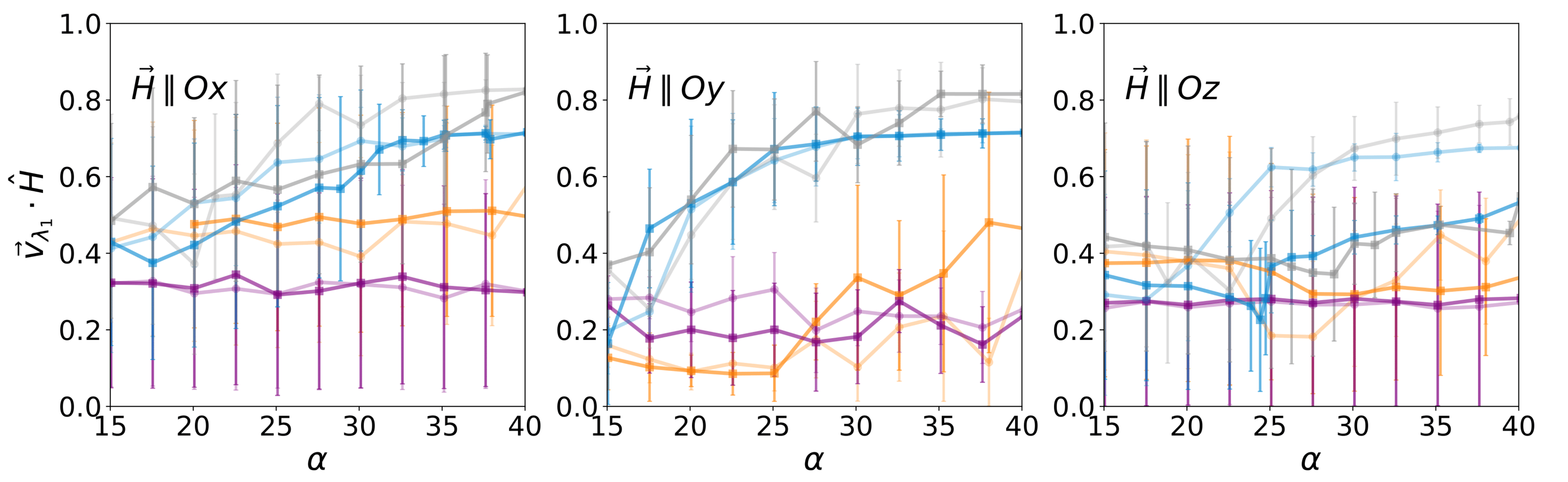}}
    \hfill
         \subfigure[]{\label{fig:all-flow-orient}\includegraphics[width=0.7\linewidth]{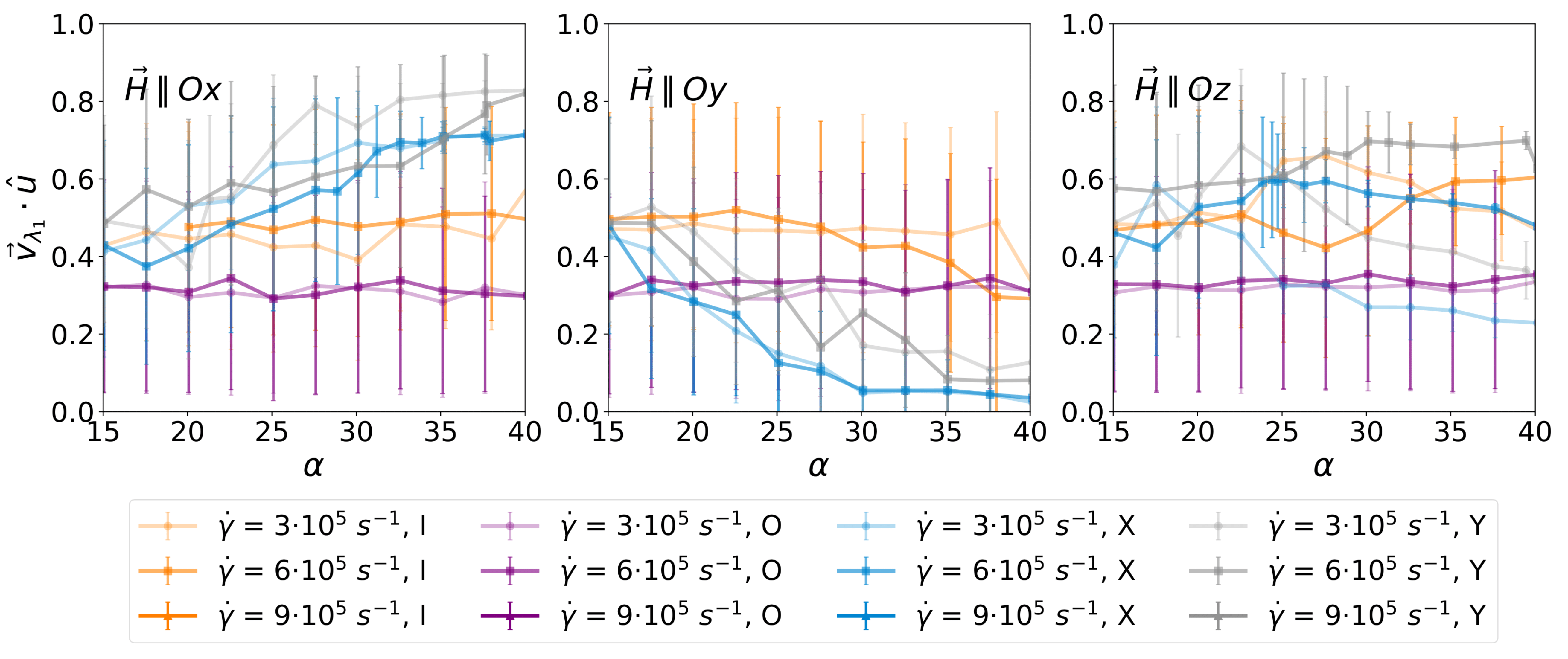}}
    \hfill
    \caption{(a) Asphericity, $Q(\alpha)$. (b) The scalar product between the eigen vector, $\vec{v}_{\lambda_1}$, of a cluster gyration tensor corresponding to the largest eigen value, $\lambda_1$, and a normalised applied magnetic field, $\hat{H}$. (c) The scalar product of $\vec{v}_{\lambda_1}$ and the normalised flow velocity, $\hat{u}$. Three panels correspond to different orientations of the applied field as shown in their bodies from the left to the right:  the field is parallel to the flow velocity, $\vec{H}  \parallel  Ox$; the field is perpendicular both to the flow gradient and to the flow velocity, $\vec{H}  \parallel  Oy$; the field along the flow gradient, $\vec{H} \parallel Oz$. SSMP topologies are differentiated by colours -- blue for X-like, grey for Y-like, orange for linear and purple for rings, the curves for three different shear rates, $\dot{\gamma}$ are plotted as detailed in the legend below the plots.}\label{fig:q-orient}
\end{figure*}

For all three orientations of $\vec{H}$, the radius of gyration for clusters formed by Y-  and X-like SSMPs grows with $\alpha$ (grey and blue curves) and this growth is not strongly affected by the  change of the shear rate. The values of the latter are provided in the legend below the plots and the more intensive the colour the higher $\dot{\gamma}$.  The increase of $ R_g (\alpha)$ is particularly pronounced for the right-most graph, where an applied field is parallel to the vertical axis, $\vec{H} \parallel  Oz$. In this orientation, the field is coaligned with the flow velocity gradient. Only under this condition, one can observe a slight growth of the $R_g(\alpha)$ for chain-like SSMPs (denoted with I, orange curves), particularly, for the highest value of $\dot{\gamma}$. For the other two mutual orientation of the flow velocity gradient and an applied magnetic field, we see that $R_g(\alpha)$ for the clusters made of chain-like SSMPs stays basically constant. Similarly, no impact of the magnetic field or shear rate on the gyration radius of clusters formed by ring-like SSMPs is observed (purple curves) as it can be seen in SupVideo2. Taking a closer look at asphericity, calculated as 

\begin{equation}
 Q (\alpha) =  \frac{3}{2} \lambda^2 - \frac{R^2_g(\alpha)}{2} ,
 \label{eq:asph}
\end{equation}

\noindent where $\lambda$ is the largest eigen value of the cluster gyration tensor, one can see that clusters made of X- and Y-like SSMPs (blue and grey curves) deform rather strongly with growing field intensity, see Fig. \ref{fig:qh}. As for the clusters made of chains, only high field parallel to the flow velocity gradient can lead to the alternation of a spherical shape (right most graph, $\alpha>30$).

It is worth noting, that in the absence of the flow, the field, required to break the clusters made of chain-like SSMPs, corresponds to $\alpha\sim 30$, while for X- and Y-like this field is found to be almost thirty per cent lower \cite{zverev2021,novak2019}. This evidences that the flow  stabilises the spherical shape of the SSMP clusters.  

One can calculate the cluster principle axis as an eigen vector, $\vec{v}_{\lambda_1}$, of the cluster gyration tensor that corresponds to the largest eigen value, denoted by $\lambda_1$. In Fig. \ref{fig:all-field-orient}, we plot the absolute value of the cosine of the angle between the orientation of $\vec{H}/|\vec{H}| = \hat{H}$ and $\vec{v}_{\lambda_1}$. The closer the value of the cosine  to unity, the more the principle axis of the cluster is coaligned with the field. In case the field and the flow velocity are parallel (left most plots in Figs. \ref{fig:qh} and \ref{fig:all-field-orient}), clusters made of Y- and X-like SSMPs are nearly equally elongated at a given field and shear rate. Both clusters tend to decrease the angle between their main axis with the flow-field direction as the field grows. At the highest value of $\alpha$ they assume the angle of approximately 45$^{\circ}$ with $\vec{H}$ as cosine approaches 0.7. In this configuration of flow and field, neither clusters made of rings, nor those of chains, acquire any pronounced asphericity independently from the value of the Langevin parameter, $\alpha$. So, it is not surprising that their orientation in the flow is not affected by the field strength (see, purple and orange curves in Fig. \ref{fig:all-field-orient}).

The same can be commented about the middle plot of Fig. \ref{fig:all-field-orient}, $\vec{H} \parallel Oy$: the clusters made of chain- and ring-like SSMPs are almost spherical for any value of $\alpha$, and are hardly exhibiting any dependence of the main axis orientation on the applied field. For this geometry, clusters made of Y- and X- SSMPs align their main axes at 45$^{\circ}$ with the field direction already for $\alpha\sim20$. At the same time, the angle between the cluster main axis and the fluid velocity, $\hat{u}=\vec{u}/|\vec{u}|$, which cosine is plotted in Fig. \ref{fig:all-flow-orient}, goes to 90$^{\circ}$ with growing field (middle plot in Fig. \ref{fig:all-flow-orient}, blue and gray curves). This means that the clusters made of Y- and X-like SSMPs assume an angle of approximately 40 $^{\circ}$ with the field and align perpendicular to the flow.  If $\vec{H} \parallel Oz$, for high values of $\alpha$,  the clusters made of chain-like SSMPs get aspherical, but the orientation of their main axes is found to be field-independent and is characterised by approximately 60$^{\circ}$ with fluid velocity vector (Fig. \ref{fig:all-flow-orient}, right most, orange curves), while the angle with the field is a little larger, $\sim 70^{\circ}$. In general, in this geometry, only the orientation of clusters made of X- and Y-like SSMPs for the lowest shear rate is affected by the growth of $\alpha$. It is worth mentioning that the large error-bars obtained for the plots in Fig. \ref{fig:q-orient} are related to averaging over oscillating data, thus, the averages should be considered with caution as they only reflect the mean values, while the amplitudes are not reflected.

\begin{figure}[h!]
    \centering
    \subfigure[]{\label{fig:stretched-X-Hy}
     \includegraphics[width=0.49\linewidth]{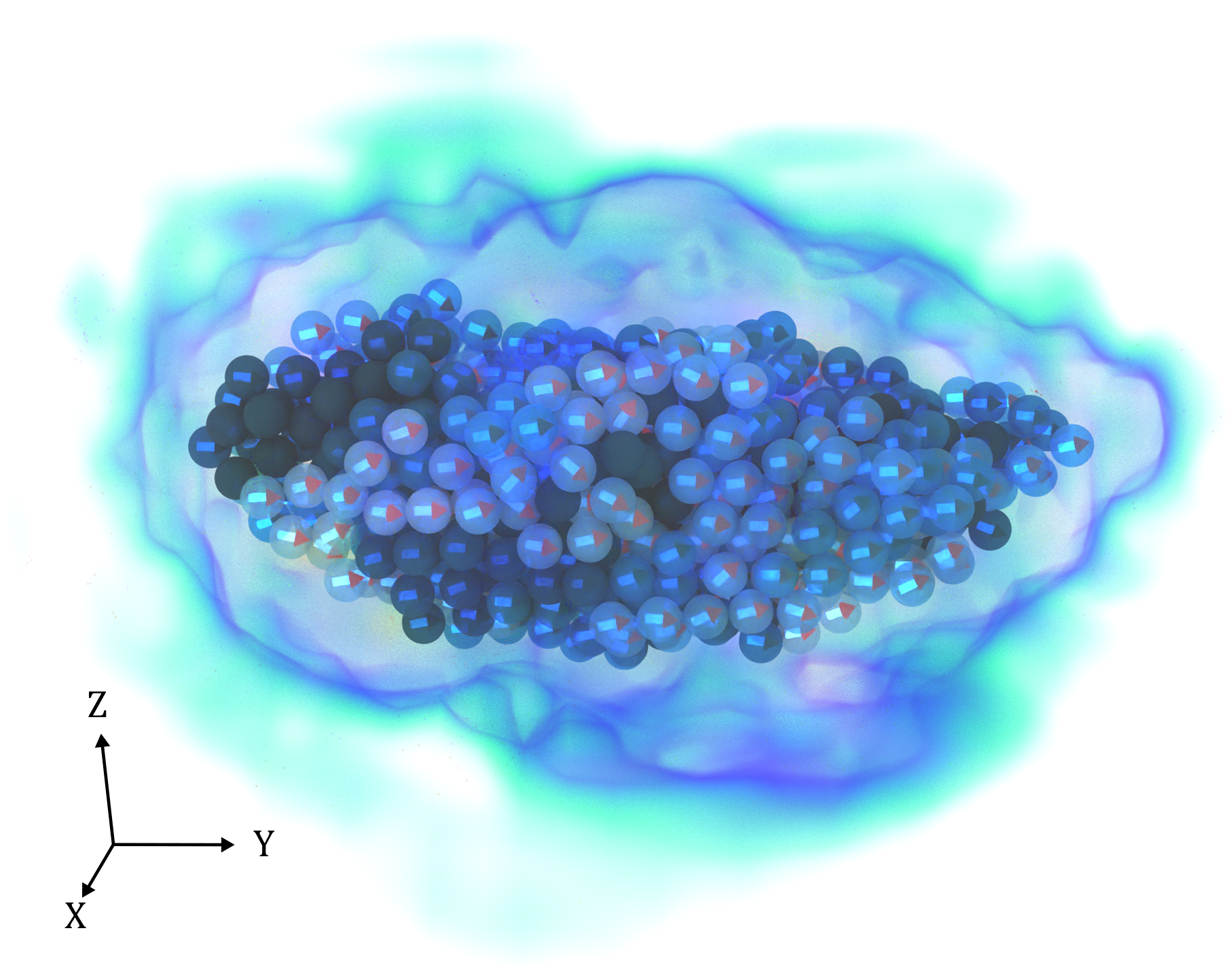}}
    \subfigure[]{\label{fig:stretched-Y-Hz} \includegraphics[width=0.39\linewidth]{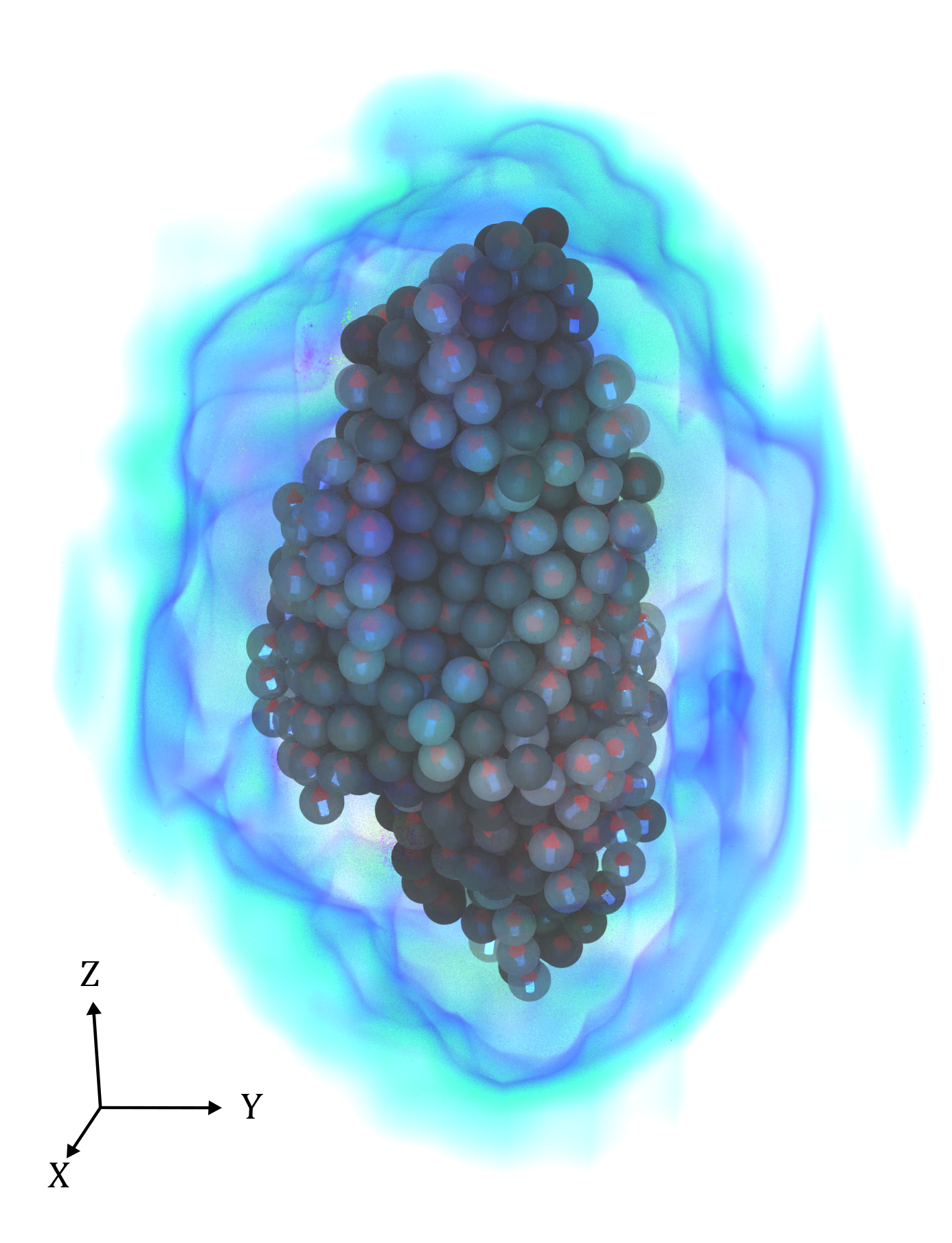}}
    \hfill
    \caption{Typical snapshots of differently elongated clusters: (a) cluster made of X-like SSMPs;  $\alpha = 42.67$; $\dot{\gamma} = 3 \cdot 10^5 s^{-1}$; $\vec{H}  \parallel  Oy$; (b) cluster made of Y-like SSMPs; $\alpha = 42.36$; $\dot{\gamma} = 3 \cdot 10^5 s^{-1}$; $\vec{H} \parallel Oz$.  Effective hydrodynamic fields around clusters are shown as halos, shear and thermal components were filtered out.}
\end{figure}

In order to visualise the alignment, in Figs. \ref{fig:stretched-X-Hy} and \ref{fig:stretched-Y-Hz}, we present characteristic snapshots of the clusters made of respectively X- and Y-like SSMPs.  On the left, in Fig. \ref{fig:stretched-X-Hy}, in which $\vec{H} || Oy$, $\alpha\sim 42$, one sees a typical rather strong elongation of the cluster along $y$-axis. Similar degree of the alignment is observed on the right, Fig. \ref{fig:stretched-Y-Hz}, in which a field of a similar strength is pointing along $Oz$. Here we show the flow fields created by the motion of the clusters. All points that have the same velocity modulus are assigned a colour: the closer to the cluster, the higher the velocity. Note that thermal noise and shear are subtracted from the fluid velocities and the absolute values are irrelevant. The halos for both clusters look very similar as expected from the analysis above.  

Overall, independently from the field orientation, the flow clearly stabilises clusters made of linear SSMPs and in comparison to a flow-free case, those clusters basically do not deform in the field. It can be attributed to the fact that the flow assists the rearrangements of chains inside the cluster, so that they can optimise both Zeeman and dipolar interactions without a significant penalty on the steric attraction. Due to precrosslinking that is not favourable for the dipole-field alignment in the case of Y- and X-like SSMPs, such clusters cannot be fully stabilised in their spherical shape by the flow, but the disintegration is clearly prevented. The direction of the elongation of such clusters is defined by the strength of an applied field. For high fields and low shear rates  the angle between the main axis and the field approaches 40-45$^{\circ}$.  

Concluding this section, one can underline the overall promising cargo properties of the clusters made of SSMPs. Even though the deformations are the weakest for the case of clusters made of linear and ring-like SSMPs, it is too early to conclude if those are optimal for the drug targeting, as their responsiveness might leave much to be desired. In order to investigate the latter, in the next section, we discuss the time evolution of clusters shape.
\begin{figure*}[!ht]
 \centering
 \subfigure[]{\label{fig:outlier_dom_freq_I_Ox} \includegraphics[width=0.32\textwidth]{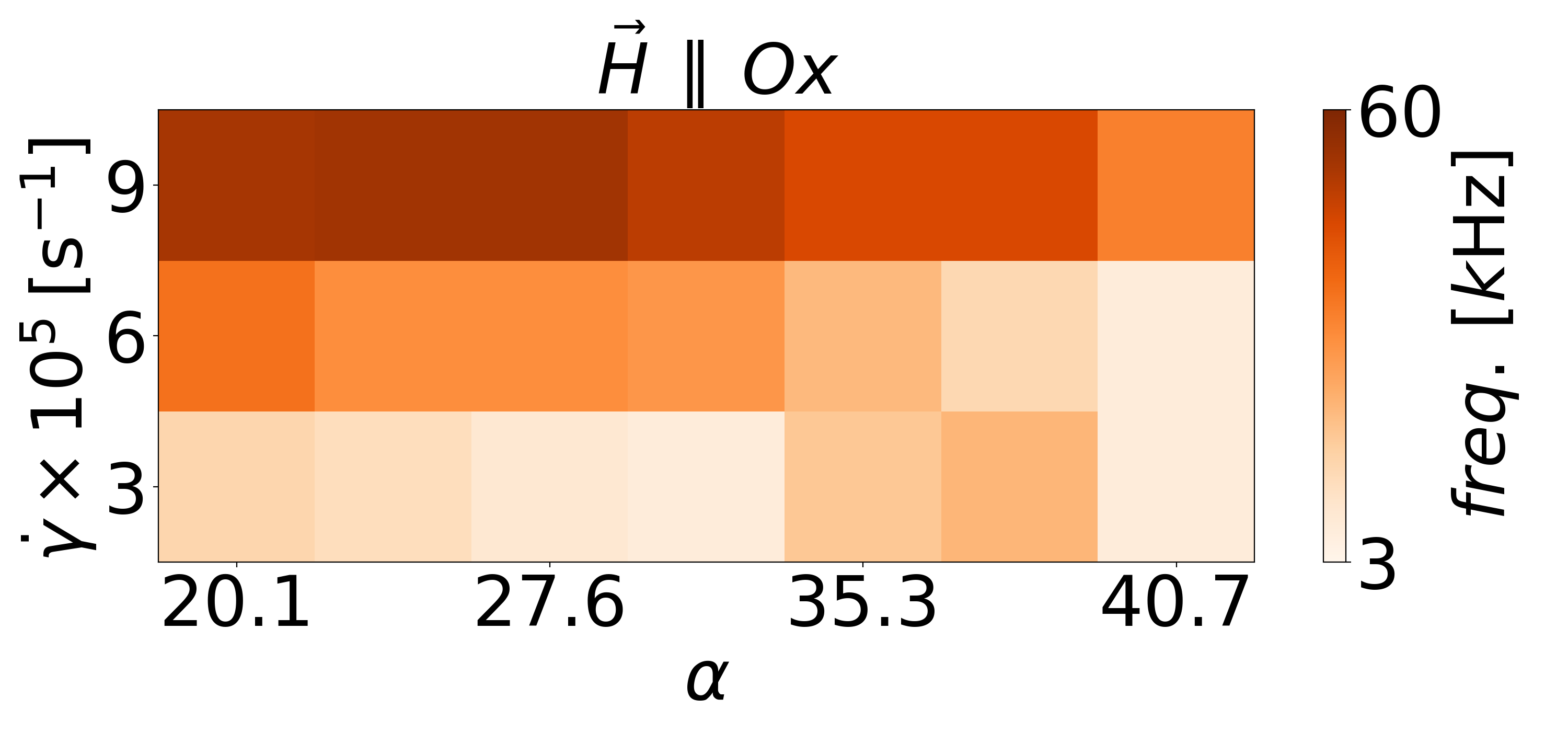}}
    \hfill
 \subfigure[]{\label{fig:outlier_dom_freq_I_Oy} \includegraphics[width=0.32\textwidth]{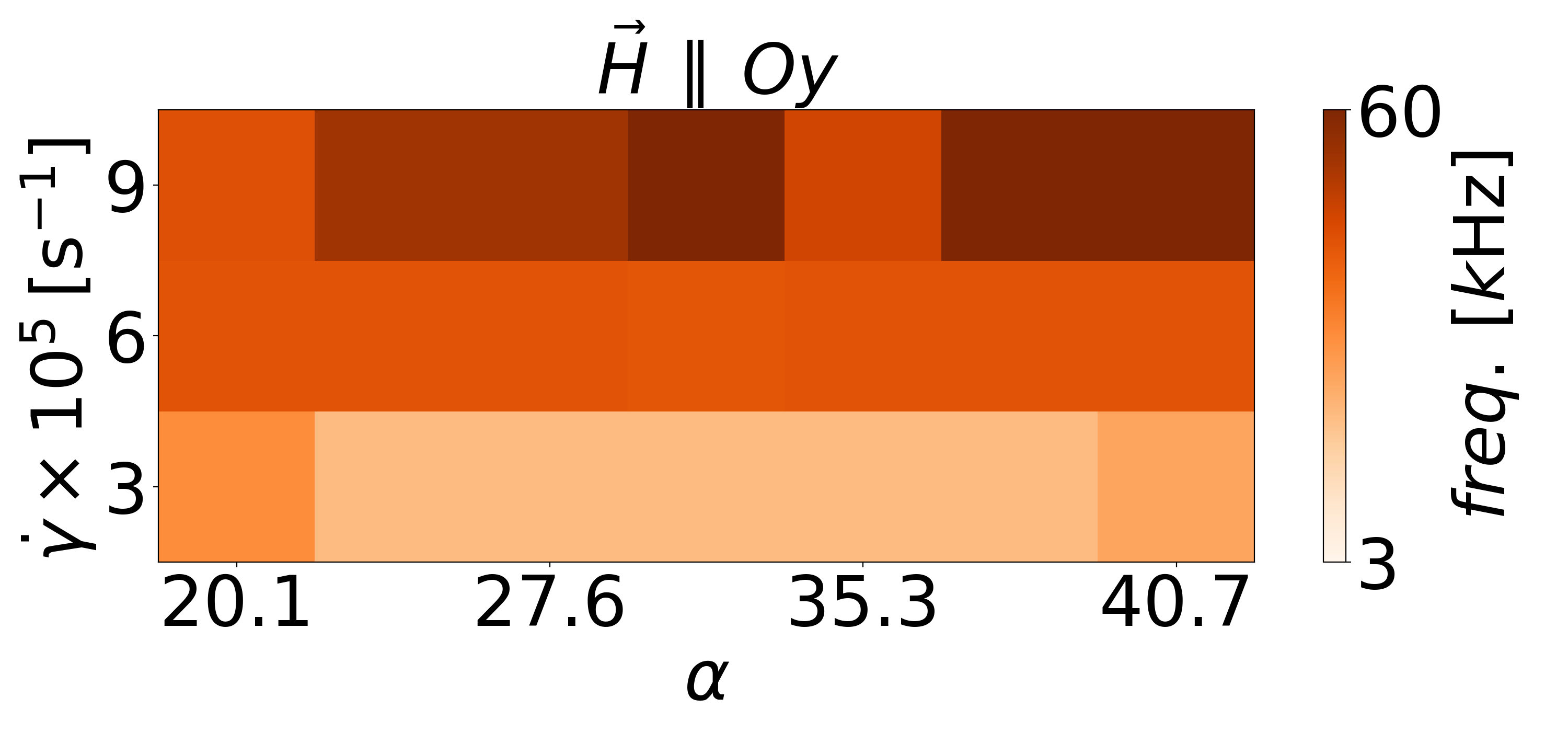}}
    \hfill
 \subfigure[]{\label{fig:outlier_dom_freq_I_Oz}\includegraphics[width=0.32\textwidth]{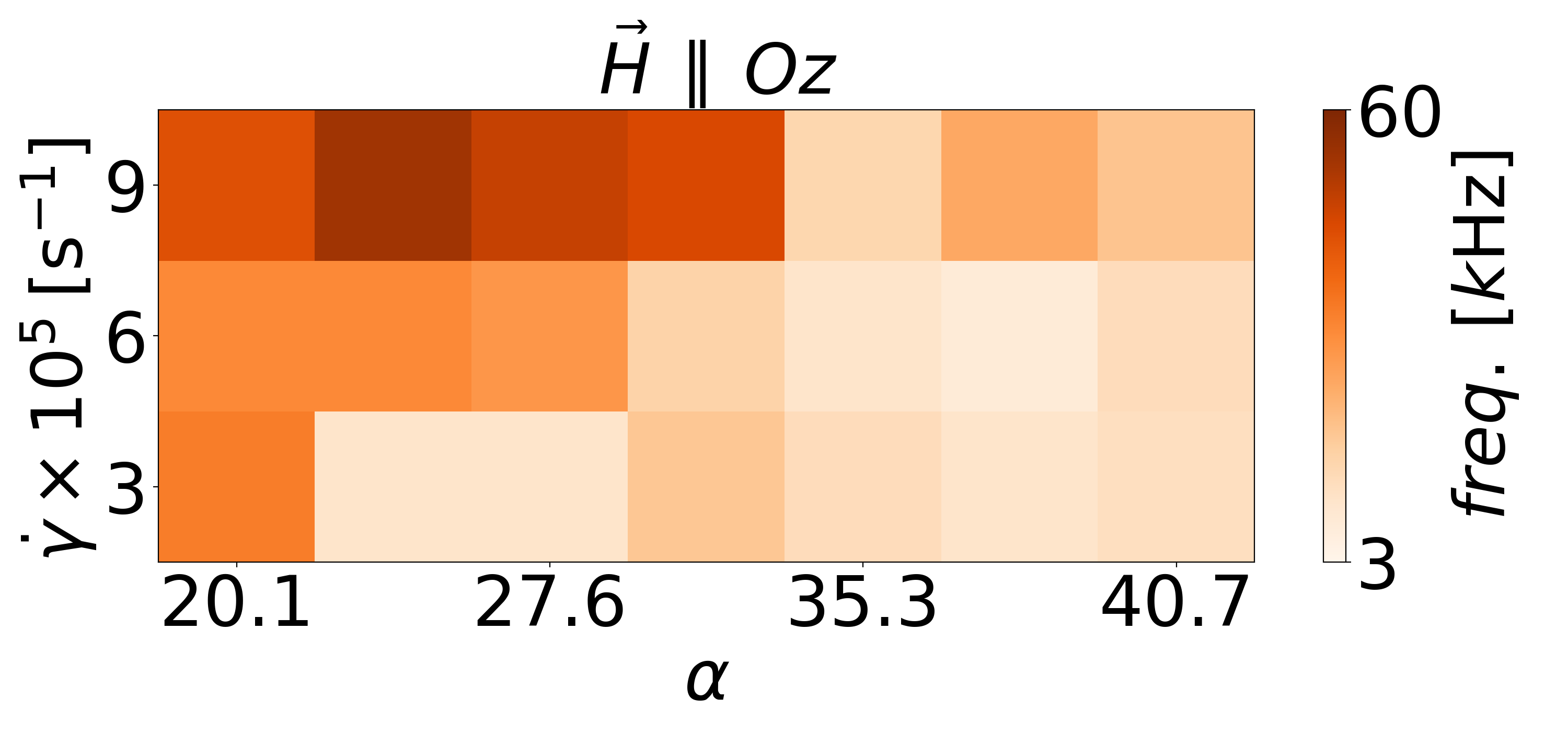}
       }
    \hfill
\subfigure[]{\label{fig:outlier_dom_freq_Y_Ox}\includegraphics[width=0.32\textwidth]{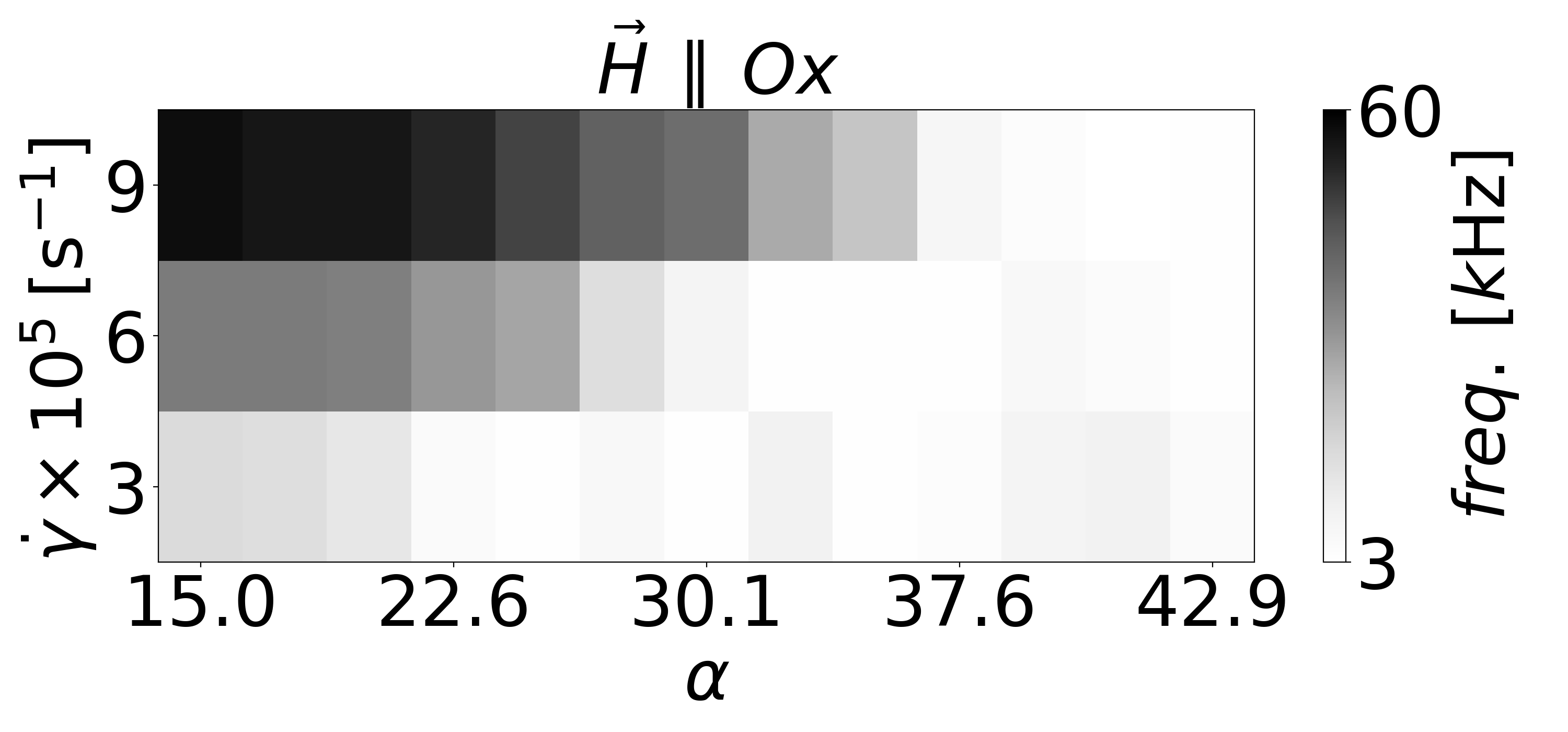}}
    \hfill
 \subfigure[]{\label{fig:outlier_dom_freq_Y_Oy}\includegraphics[width=0.32\textwidth]{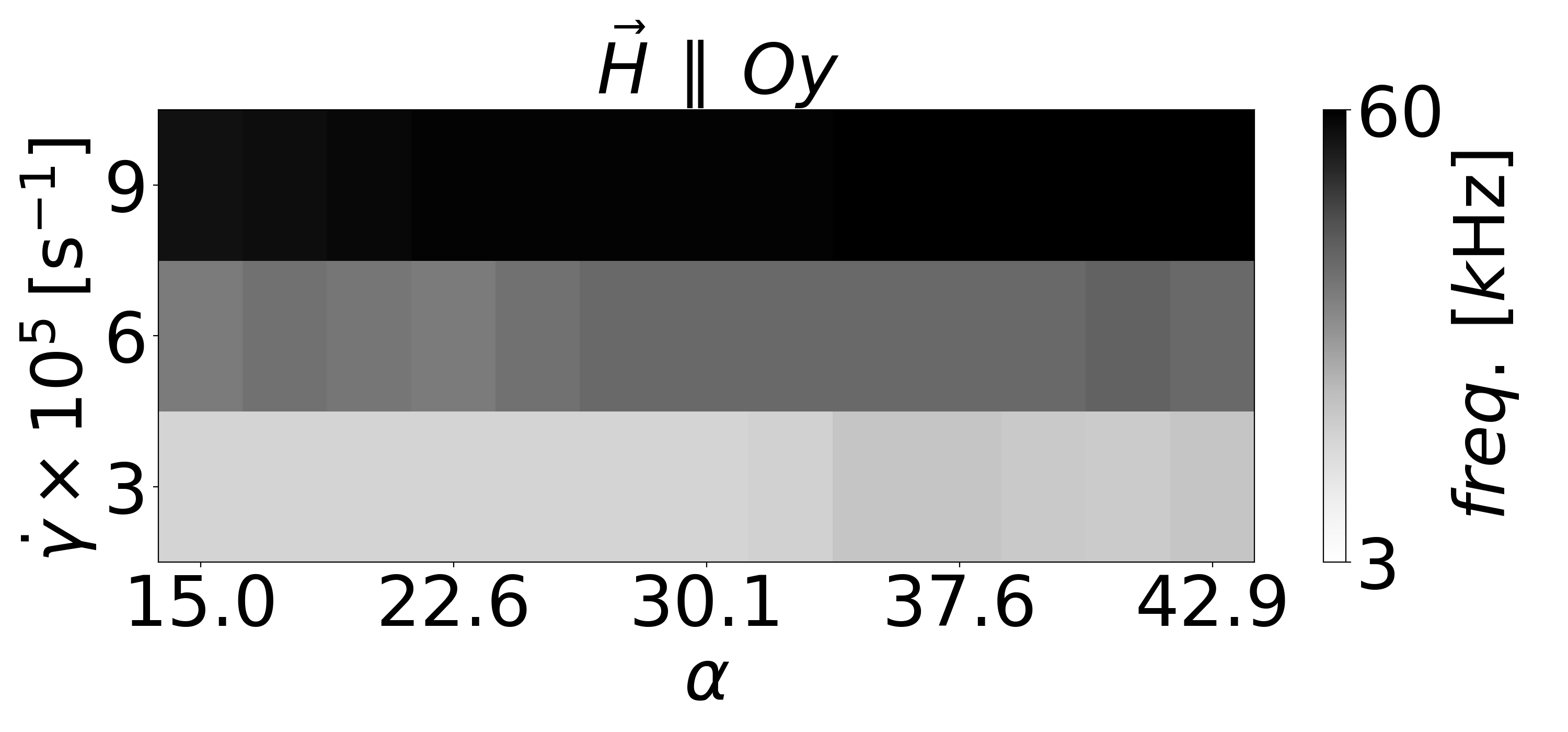}}
    \hfill
 \subfigure[]{\label{fig:outlier_dom_freq_Y_Oz}\includegraphics[width=0.32\textwidth]{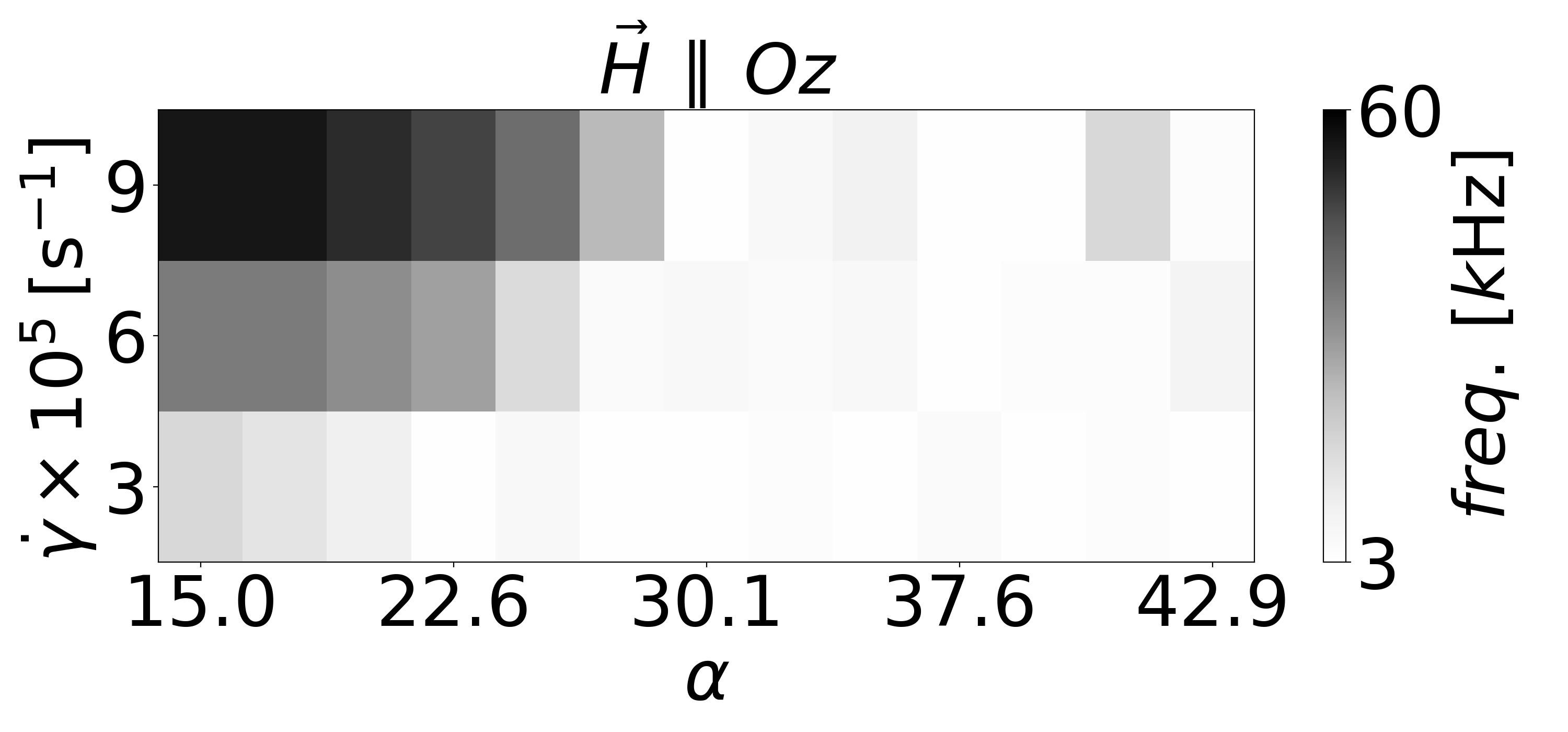}}
     \hfill
\subfigure[]{\label{fig:outlier_dom_freq_X_Ox} \includegraphics[width=0.32\textwidth]{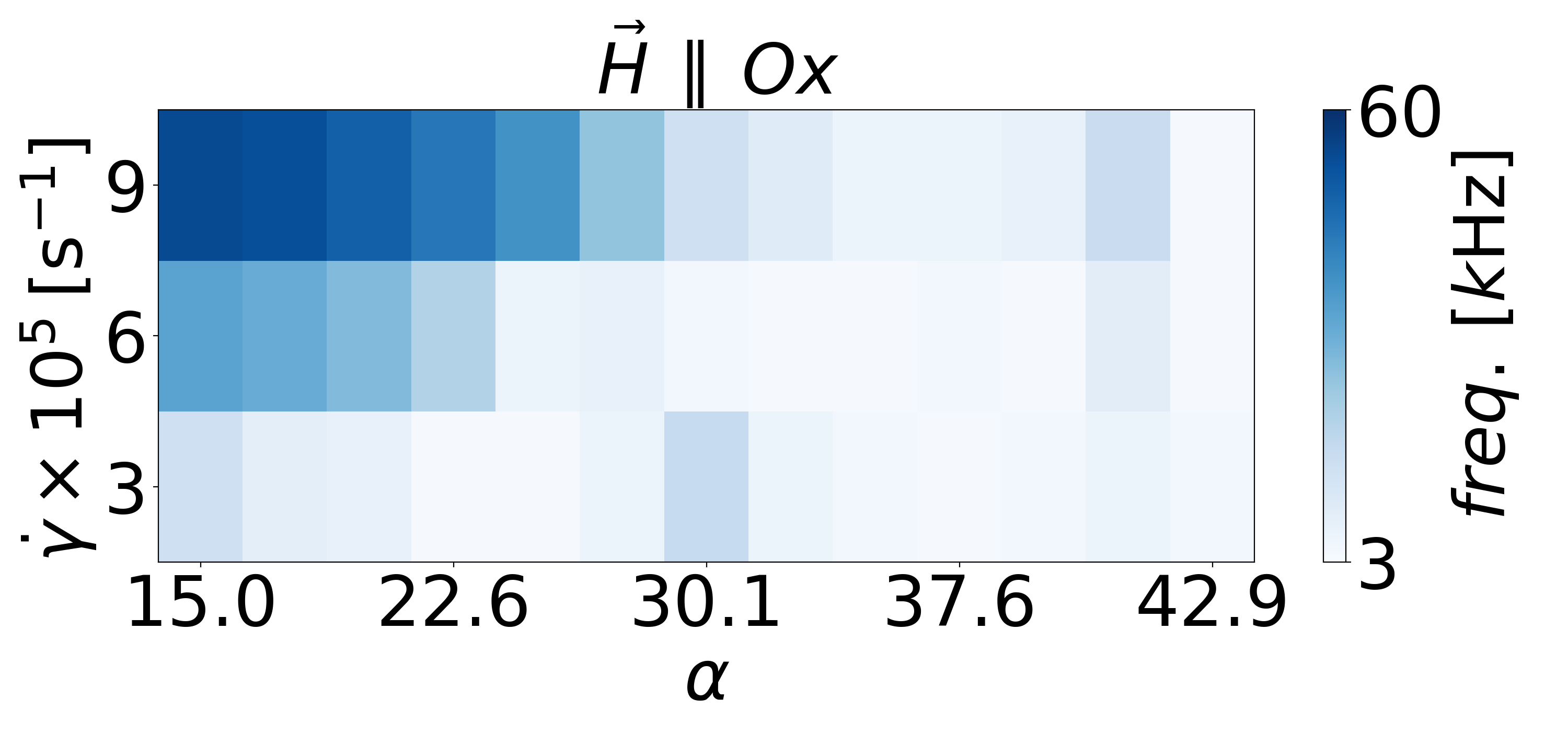}}
    \hfill
 \subfigure[]{ \label{fig:outlier_dom_freq_X_Oy}\includegraphics[width=0.32\textwidth]{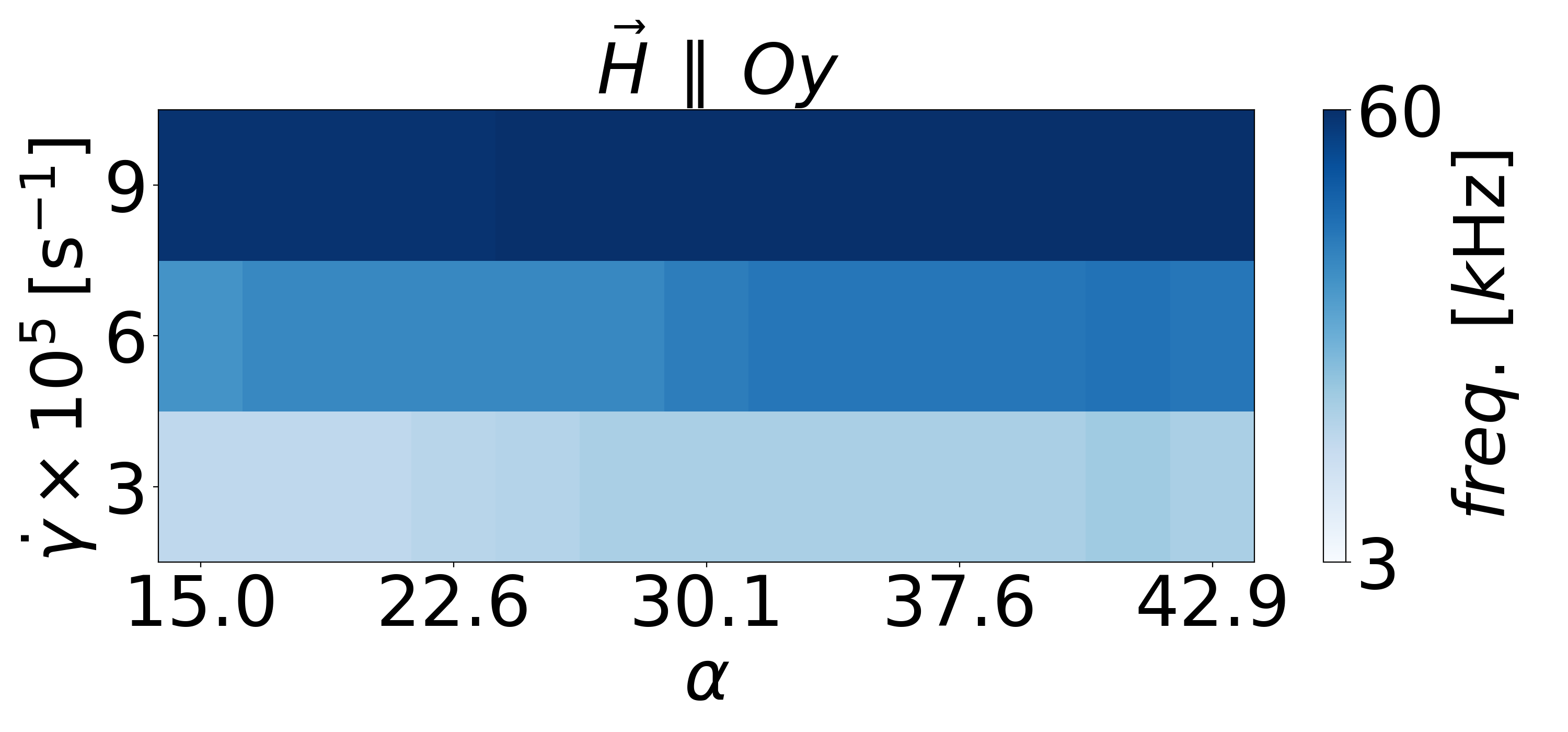}}
    \hfill
 \subfigure[]{\label{fig:outlier_dom_freq_X_Oz}\includegraphics[width=0.32\textwidth]{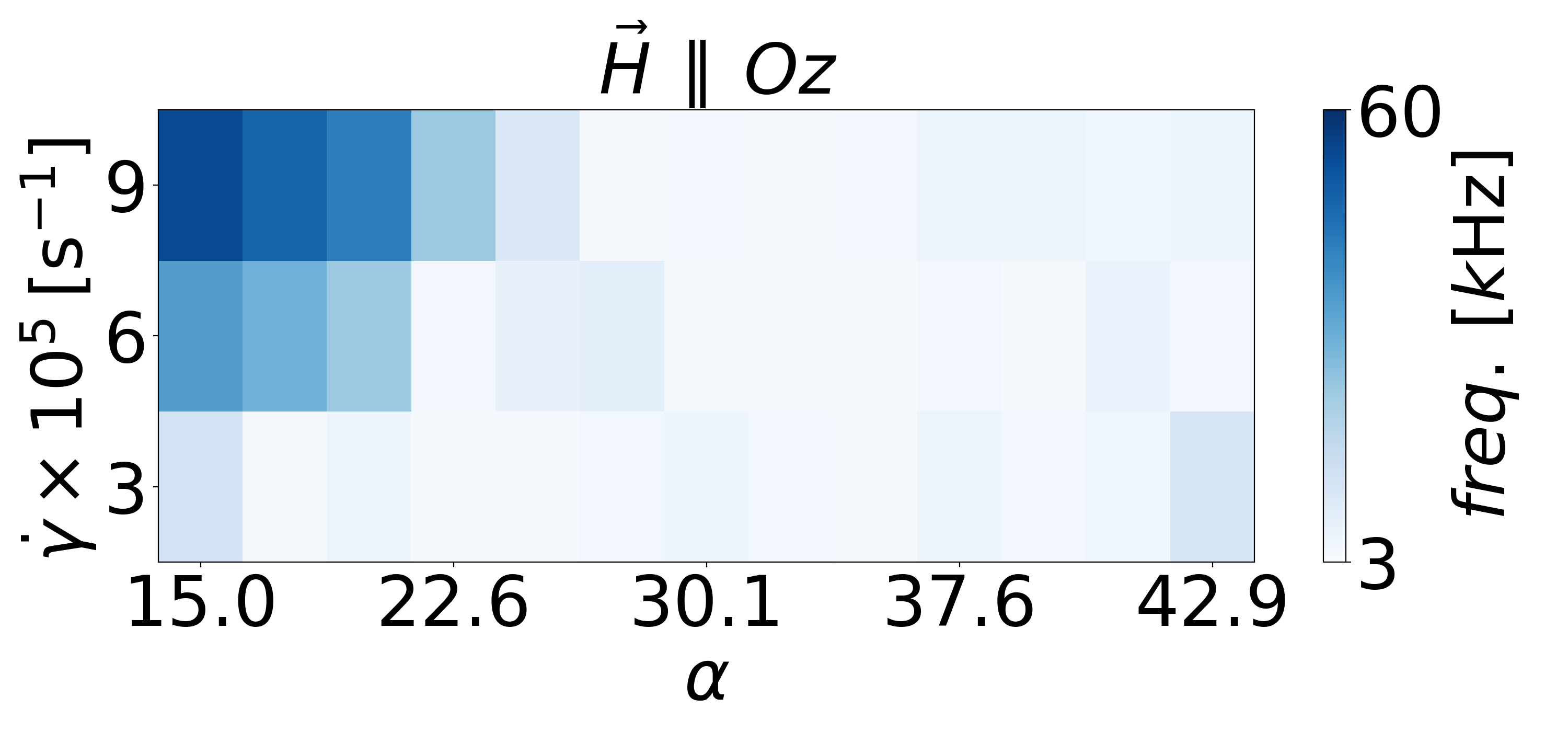}}
    \hfill
\subfigure[]{ \label{fig:outlier_dom_freq_Rings_Ox}\includegraphics[width=0.32\textwidth]{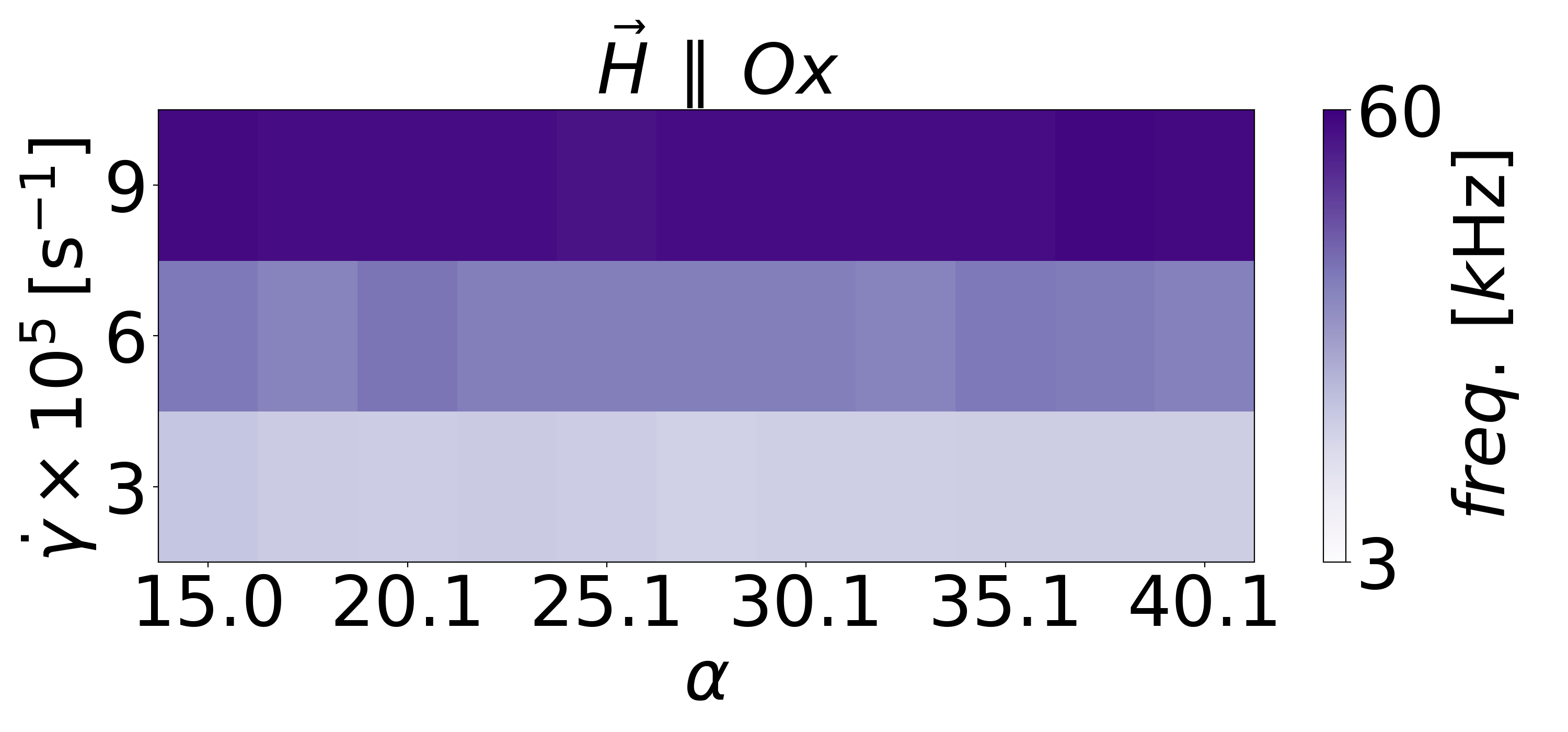}}
    \hfill
 \subfigure[]{\label{fig:outlier_dom_freq_Rings_Oy}\includegraphics[width=0.32\textwidth]{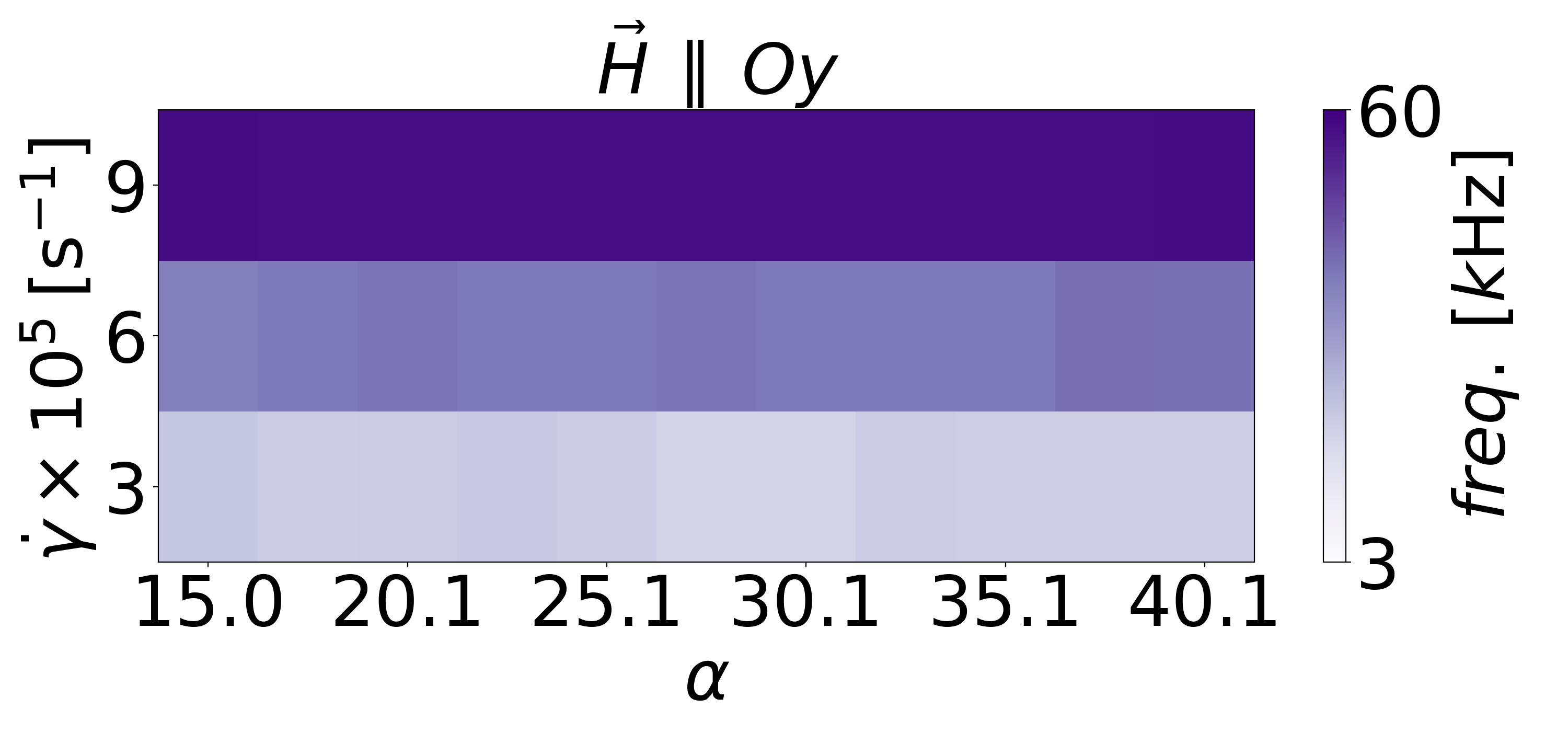}}
    \hfill
 \subfigure[]{\label{fig:outlier_dom_freq_Rings_Oz}\includegraphics[width=0.32\textwidth]{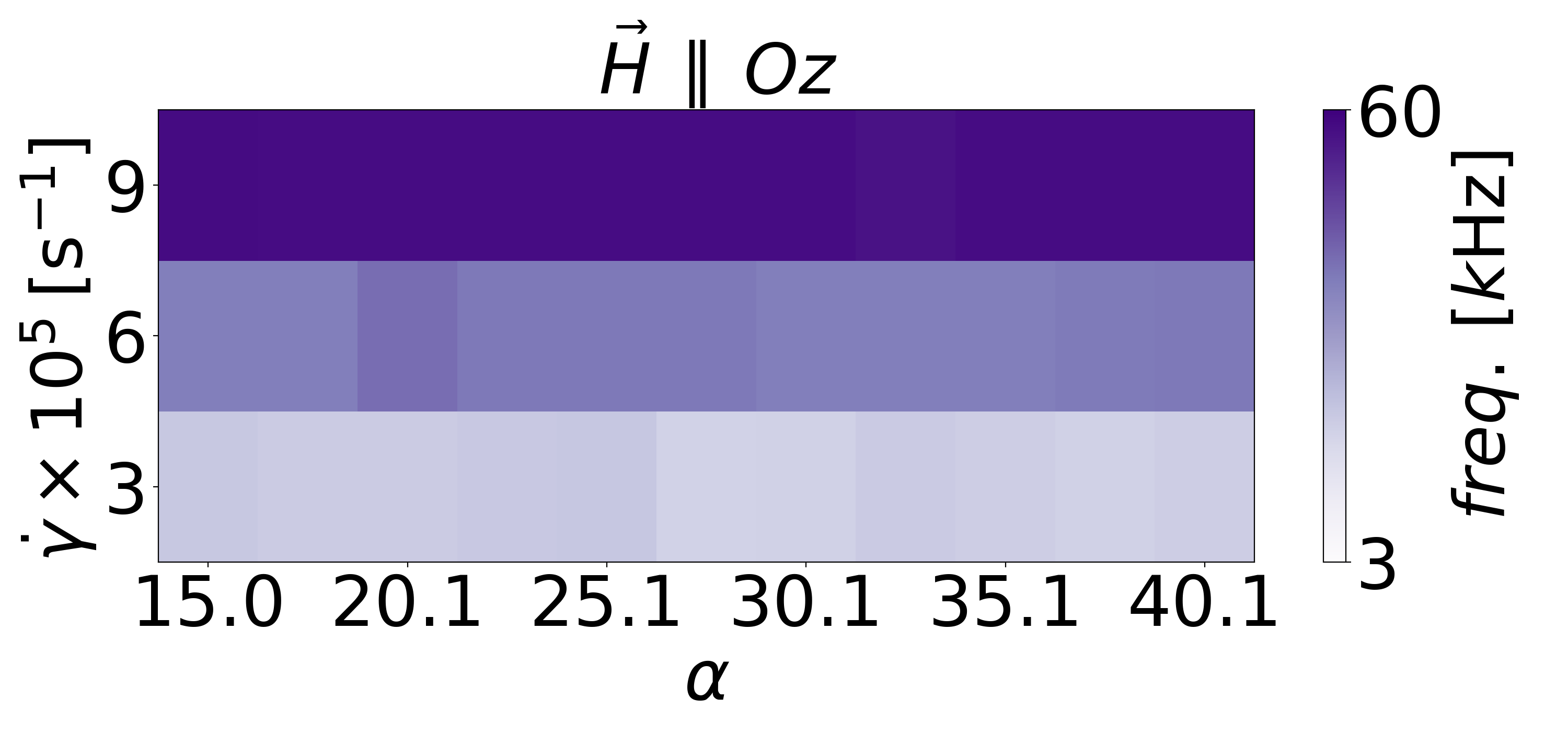}}
    \hfill
        \caption{
        Heatmaps of cluster rotational frequencies around $y$-axis in log scale for a specific topology of the SSMPs forming the cluster: (a) - (c) chains, (d) - (f) X-like, (g) - (i) - Y-like, and, finally, (j) - (l)ring-like ones. Columns correspond to a given orientation of the applied field.
        }
       \label{fig:rot-t}
  \end{figure*}

 \subsection{Cluster Dynamics}\label{subs:def-time}

One needs to distinguish between two processes happening to the clusters under the combined action of the field and the shear: first is the tumbling, second is the wobbling or shape deformations. Tumbling can be described as follows. Let us fix a particle on the surface of a cluster. If we connect the centre of this particle and the centre of mass of the cluster, we obtain a vector, whose angular velocity characterises the rotational behaviour of the cluster in a shear flow.  We chose the $z$-component of the latter vector, calculated its time-dependencies for all orientations of $H$ and SSMP topologies. Next, we Fourier-transformed the results and obtained typical rotation frequencies. The results are presented in Fig. \ref{fig:rot-t} in a form of heatmaps. Here, one can distinguish two types of behaviour. In Figs. \ref{fig:outlier_dom_freq_I_Ox} -- \ref{fig:outlier_dom_freq_Y_Oy}, \ref{fig:outlier_dom_freq_Y_Oz}, \ref{fig:outlier_dom_freq_X_Ox}, \ref{fig:outlier_dom_freq_X_Oz} the rotational frequency is not only $\dot{\gamma}$-, but also $H$-dependent with the highest values concentrated in a left upper corner. The rotations are clearly blocked by high fields aligned along $Ox$ and $Oz$. If the field is parallel to $Oy$, its influence on the rotational frequency is basically negligible. It is particularly well represented by Figs. \ref{fig:outlier_dom_freq_Y_Oy}, \ref{fig:outlier_dom_freq_X_Oy} and \ref{fig:outlier_dom_freq_Rings_Oy} that are for clusters made of Y-, X- and ring-like SSMPs respectively. The latter (clusters formed by ring-like SSMPs) are, in general, not affected by the application of the magnetic field. 

\begin{figure*}[h!]
 \centering
 \subfigure[]{\label{fig:c-like-qt-ox} \includegraphics[width=0.26\textwidth]{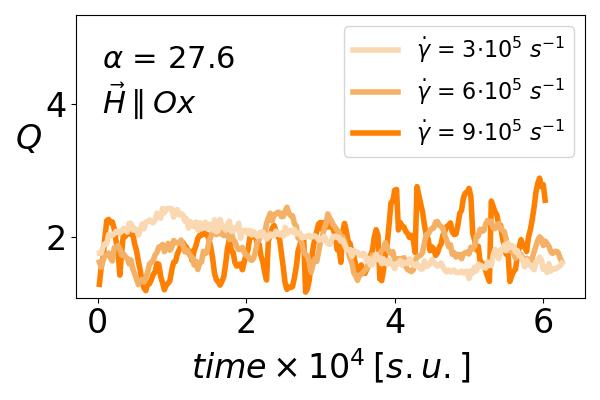}}
    \hfill
 \subfigure[]{\label{fig:c-like-qt-oy} \includegraphics[width=0.26\textwidth]{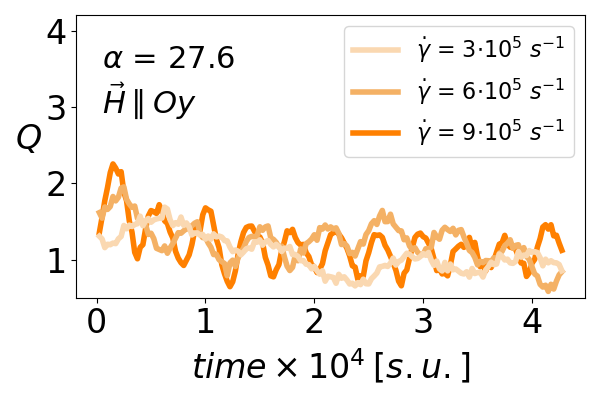}}
    \hfill
 \subfigure[]{\label{fig:c-like-qt-oz}\includegraphics[width=0.26\textwidth]{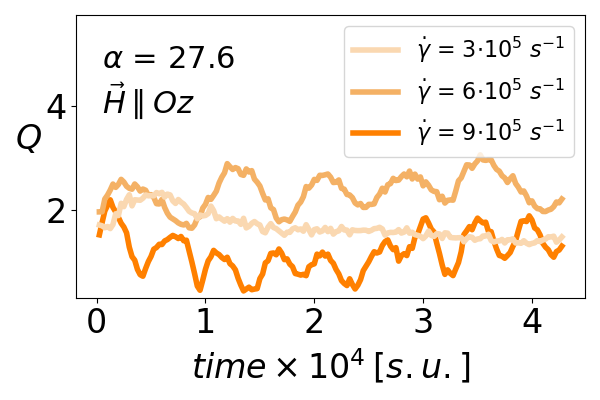}
       }
    \hfill
\subfigure[]{\label{fig:y-like-qt-ox} \includegraphics[width=0.26\textwidth]{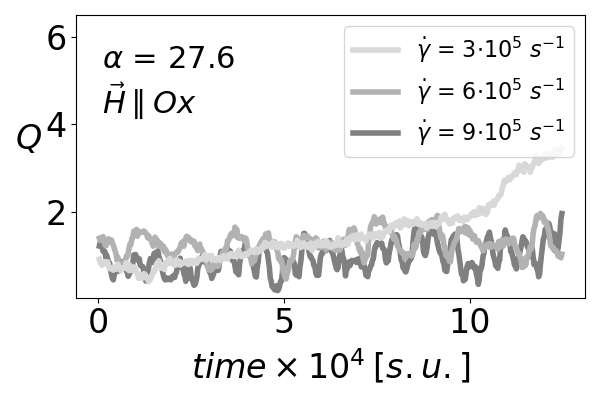}}
    \hfill
 \subfigure[]{ \label{fig:y-like-qt-oy}\includegraphics[width=0.26\textwidth]{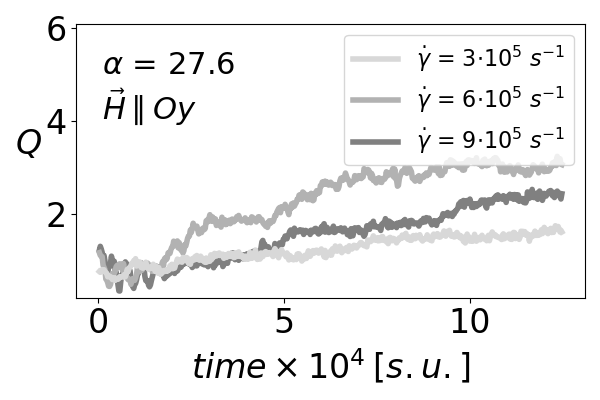}}
    \hfill
 \subfigure[]{\label{fig:y-like-qt-oz}\includegraphics[width=0.26\textwidth]{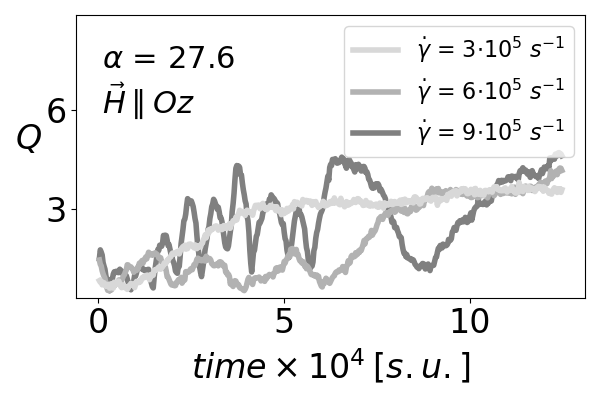}}
    \hfill
\subfigure[]{\label{fig:x-like-qt-ox}\includegraphics[width=0.26\textwidth]{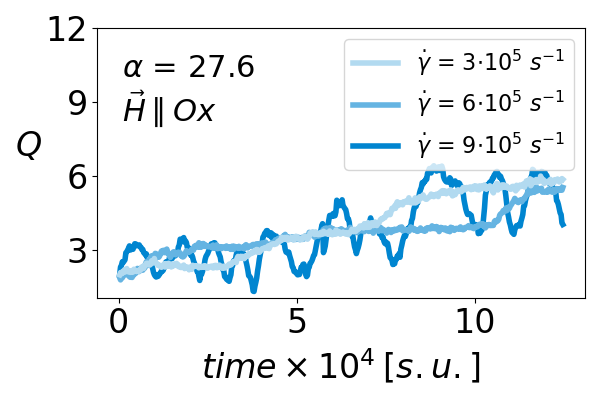}}
    \hfill
 \subfigure[]{\label{fig:x-like-qt-oy}\includegraphics[width=0.26\textwidth]{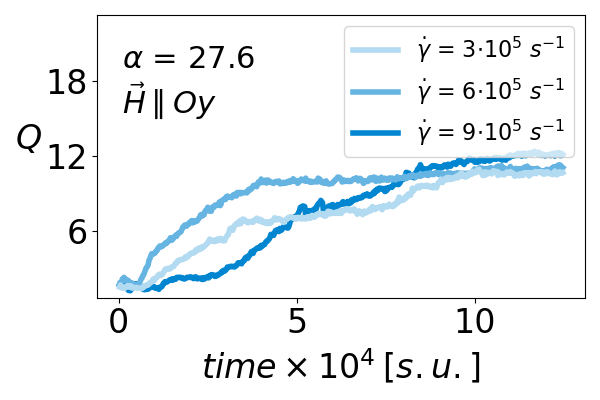}}
    \hfill
 \subfigure[]{\label{fig:x-like-qt-oz}\includegraphics[width=0.26\textwidth]{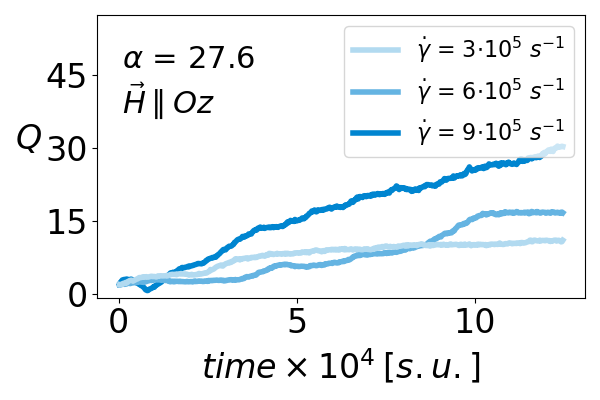}}
    \hfill
\subfigure[]{ \label{fig:r-like-qt-ox}\includegraphics[width=0.26\textwidth]{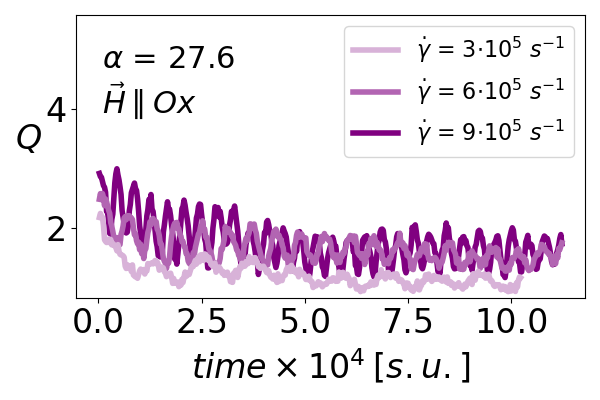}}
    \hfill
 \subfigure[]{\label{fig:r-like-qt-oy}
        \includegraphics[width=0.26\textwidth]{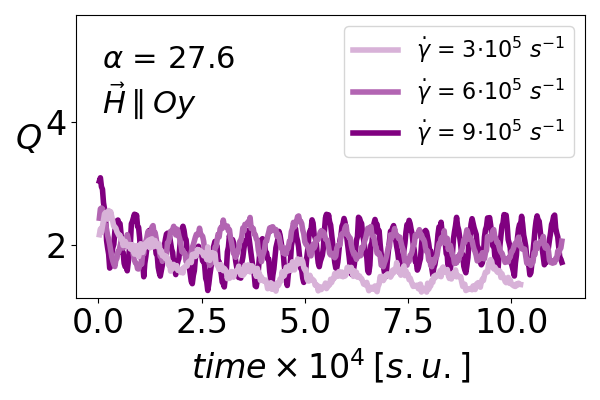}}
    \hfill
 \subfigure[]{\label{fig:r-like-qt-oz}\includegraphics[width=0.26\textwidth]{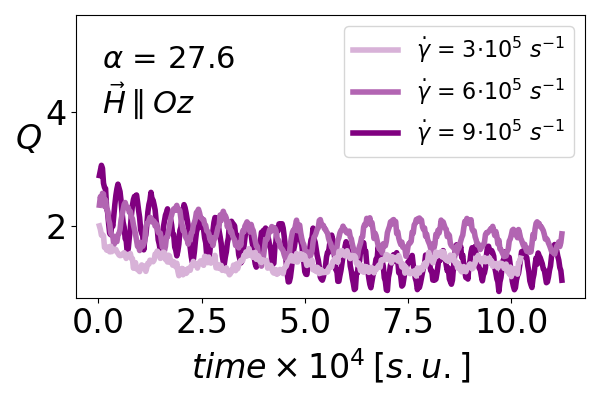}}
    \hfill
        \caption{Dynamics of asphericity $Q$ for a specific topology of the SSMPs forming the cluster: (a) - (c) chains, (d) - (f) Y-like, (g) - (i) - X-like, and, finally, (j) - (l)ring-like ones. Columns correspond to a given orientation of the applied field. The brightness of the curves corresponds to the value of the shear rate -- the darker the curve, the higher the value of $\dot{\gamma}$. The value of $\alpha=27.6$.}
       \label{fig:qt}
  \end{figure*}

The second process is wobbling that can be characterised by the time-dependence of the asphericity. The latter is presented in Fig. \ref{fig:qt}. Here, in each row we plot the dependence of $Q$ on the simulation time for a specific topology of the SSMPs forming the cluster: chains, Y-like, X-like, and, finally, ring-like ones. Columns instead correspond to a given orientation of the applied field, {\it i.e.}, in the first column the field is aligned with the flow direction, $Ox$; in the middle the field is perpendicular to both flow gradient and flow directions, $Oy$; in the right most column, the field is pointing along the flow gradient, $Oz$. The brightness of the curves correspond to the value of the shear rate -- the darker the curve, the higher the value of $\dot{\gamma}$. The value of $\alpha$, meaning the intensity of an applied magnetic field, is fixed to, $\alpha=27.59$, across the figure. If one looks at the time evolution of cluster asphericity, one can distinguish three regimes: (i) oscillations with a constant frequency which value is a growing linear function of $\dot{\gamma}$ (see, SupVideo3); (ii) non-steady oscillations caused by the competition between hydrodynamic and magnetic forces (see, SupVideo4); (iii) oscillation-free monotonic increase of $Q$ with time (see, SupVideo1). It is, however, not trivial, how to quantitatively distinguish between those. In case the field is aligned along $Ox$ or $Oz$, for certain systems we noticed that the oscillations of asphericity are approximately twice more frequent than that of tumbling. Closer look reveals that combining the frequencies of asphericity oscillations and rotations can be used as a criteria for defining the aforementioned regimes. Firstly, if tumbling and wobbling frequencies are factor of two apart, it means that the two processes are essentially one, and on each half a turn the cluster elongates. This allows us to conclude that tumbling (and as a result for some cases wobbling) frequencies have an upper limit -- that of a hard sphere with the hydrodynamic radius approximately equal to the cluster gyration radius. The modulus of the angular velocity of a sphere is known to be approximately $\dot{\gamma}/2$ \cite{lin70a} for low Re as used in this paper. In our case, only clusters made of rings have a comparable frequency, the clusters made of SSMPs with chain-, Y- and X- topologies are rotating (wobbling in some cases) much slower. If, in Fig. \ref{fig:qt} we see the noise, whose Fourier transform provides us with an artificially high frequency, we safely disregard it and not call these regimes oscillations. On the other hand, we set a lower limit for the frequency. If it is lower than that required to observe a full period of the oscillations within the simulation time, then this is not an oscillation. The aforementioned approach is used to decide if or not there are oscillations, for all but one $H||Oy$ case. In general, field-dependent rotation frequency in Fig. \ref{fig:rot-t}  is the sign of a presence of the transition between the three regimes: lightest points correspond to no oscillations (iii), darker to unsteady oscillations (ii), the darkest to steady oscillations (i).

Let us discuss Fig. \ref{fig:qt} in more detail.

The weakest magnetic field at which the oscillations get suppressed is found for clusters made of X-like SSMPs that can be associated with the highest magnetic susceptibility that those clusters exhibit in comparison to their counterparts.

Starting from the bottom row with ring-like topology of SSMPs, we see a typical droplet behaviour independent from the orientation of the applied field, Figs.  \ref{fig:r-like-qt-ox}--\ref{fig:r-like-qt-oz}. The amplitudes, the frequencies, as well as the means, here depend exclusively on the shear rate: the frequency grows linear, the amplitude decreases a little. The situation changes dramatically, if one looks at the upper row, Figs.  \ref{fig:c-like-qt-ox}--\ref{fig:c-like-qt-oz}, where the asphericity {\it versus} time is plotted for clusters formed by chain-like SSMPs.  The most clearly pronounced oscillations for all three values of shear can be seen if the field is applied along $Oy$ axis, and for the highest shear. Note, that the horizontal timescale is different between Figs. \ref{fig:r-like-qt-ox}--\ref{fig:r-like-qt-oz} and Figs. \ref{fig:c-like-qt-ox}--\ref{fig:c-like-qt-oz}, so there is no difference in frequency between the two topologies for the case  $\vec{H}||Oy$. For the other two cases, instead, if the cluster is made of chain-like aggregates the oscillations are much less regular. Although an applied magnetic field does affect the dynamics of a cluster made of chain-like aggregates, the amplitudes of the shape oscillations remain fairly small and very similar to those we discussed in Figs.  \ref{fig:r-like-qt-ox}--\ref{fig:r-like-qt-oz}. This once again shows to which extent the flow pushes together the SSMPs and as such reduces the elongation along the field. 

The same strength of an applied magnetic field results in a quite different effect if the clusters that are formed by Y- or X-like SSMPs as shown in Figs.  \ref{fig:y-like-qt-ox}--\ref{fig:y-like-qt-oz} and  \ref{fig:x-like-qt-ox}--\ref{fig:x-like-qt-oz}.  Here, while for clusters made of Y-like SSMPs, Figs.  \ref{fig:y-like-qt-ox}--\ref{fig:y-like-qt-oz}, some oscillatory regime survives for $\vec{H}||Ox$, $\vec{H}||Oz$ and high shears, for their counterparts of X-like SSMPs, the shape of the cluster keeps oscillating only in case of $\vec{H}||Ox$ and the highest value of $\dot{\gamma}$ and even for this case a steady growth of the mean value is observed with time (see, \ref{fig:x-like-qt-ox}).

An interesting effect can be observed in Fig. \ref{fig:x-like-qt-oy}. It is related to the crossover of the amplitude of $Q$. At first, high shear rate stabilises the shape of the cluster, but with time the elongation becomes more pronounced, the higher is $\dot{\gamma}$. Notice that the elongation happens in this case along the field and not along the flow as it has been previously confirmed by Fig. \ref{fig:stretched-X-Hy}.

We believe that the clusters that exhibit the least oscillations are the most reliable transport vehicles, as the periodic deformations might cause preliminary and undesired release of the cargo. To this extent, the clusters made of X- and Y-like SSMPs show the highest shape stability in the flow, albeit, overall, they deform rather strongly as we showed in the previous section. The next step in the analysis is related to clusters responsiveness to the applied magnetic fields and is provided below.

\subsection{Cluster Magnetisation}\label{subs:mag-av}
\begin{figure*}[h!]
    \centering
    \includegraphics[width=0.75\textwidth]{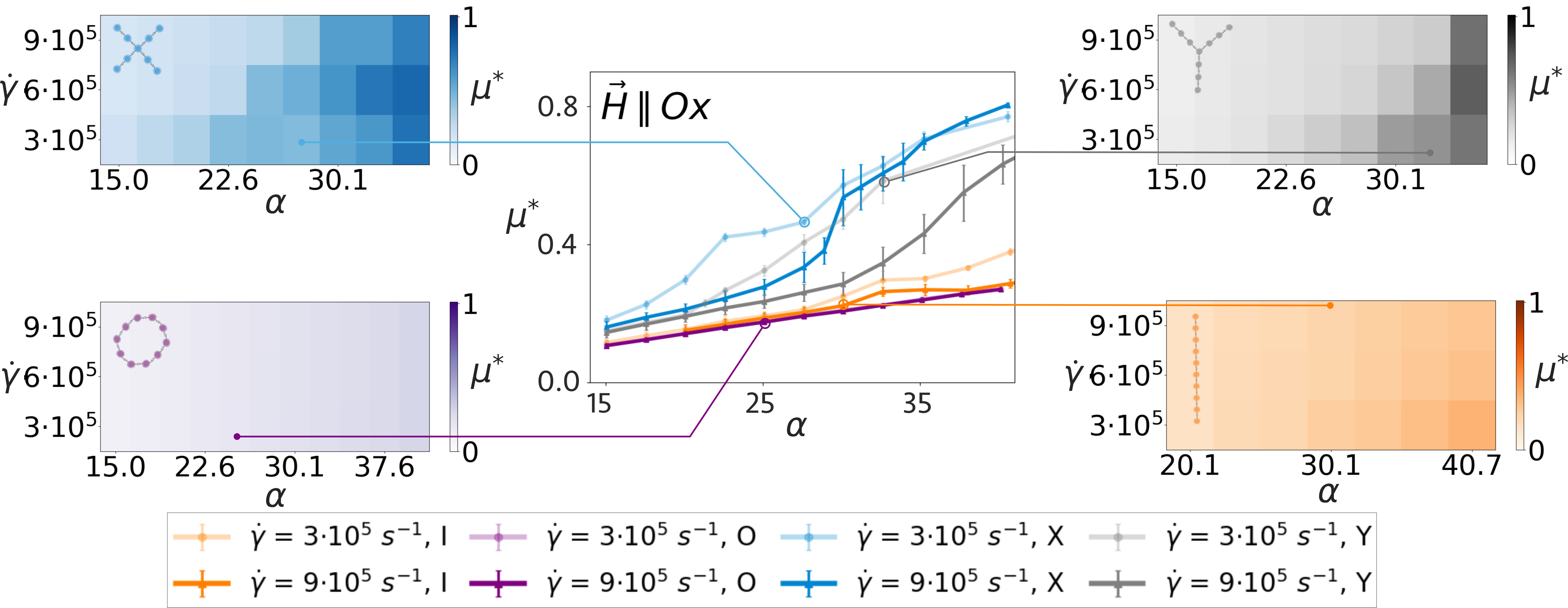}
    \caption{Magnetisation $\mu^{\ast}$ curves for all four cluster types.  The field is aligned with Ox. Middle: the total magnetic moment of the cluster, $\mu^{\ast}$, projected on the direction of an applied field and normalised by $\mu N_p$ as a function of $\alpha$. The brightness of the curves corresponds to the value of the shear rate. Exact values of $\dot{\gamma}$ are provided in the legend. Four corners: the heatmap for $\mu^{\ast}$ in axes of $\dot{\gamma}$ and $\alpha$. The darker is the colour, the higher is the value of $\mu^{\ast}$, as shown in the colour bar on the right of each diagram. }
    \label{fig:mag-vs-field-tot-ox}
\end{figure*}

Any changes of the asphericity, be that oscillations or simple elongation, cannot but affect the values of the cluster magnetisation.  In Fig. \ref{fig:mag-vs-field-tot-ox} we collect magnetisation curves for all four cluster types in the form of both curves and heat maps. 

In the middle, the plot shows how $\mu^{\ast}$ -- the total magnetic moment of the cluster, projected on the direction of an applied field and normalised by $\mu N_p$  -- depends on the value of $\alpha$. We chose here to showcase only the magnetisation for  $\vec{H}||Ox$, all the other geometries can be found in Supplementary materials.

\begin{figure}[h!]
    \centering
     \subfigure[]{\label{fig:MOx} \includegraphics[width=0.48\linewidth]{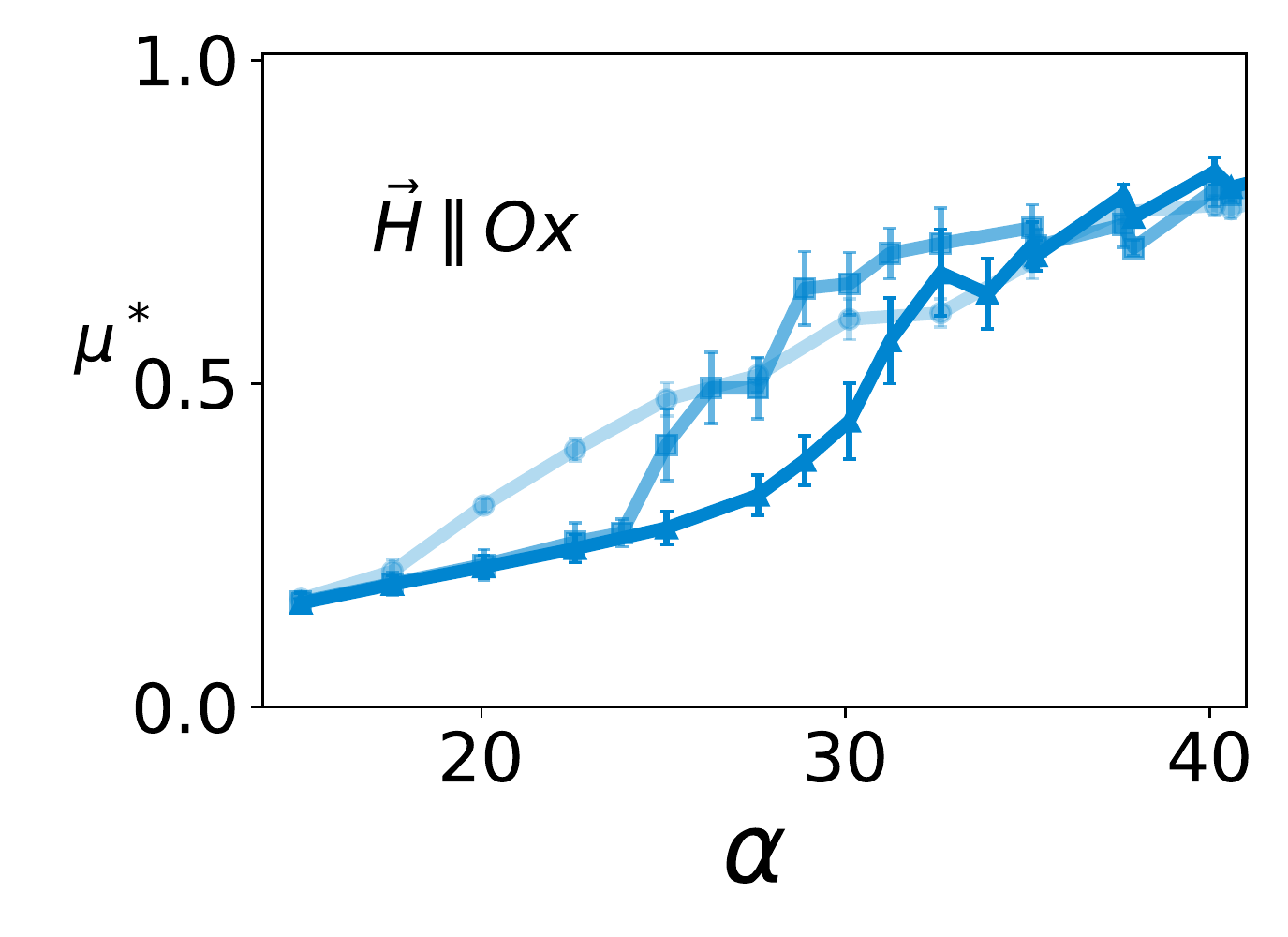}}
    \hfill
         \subfigure[]{\label{fig:MOy}\includegraphics[width=0.48\linewidth]{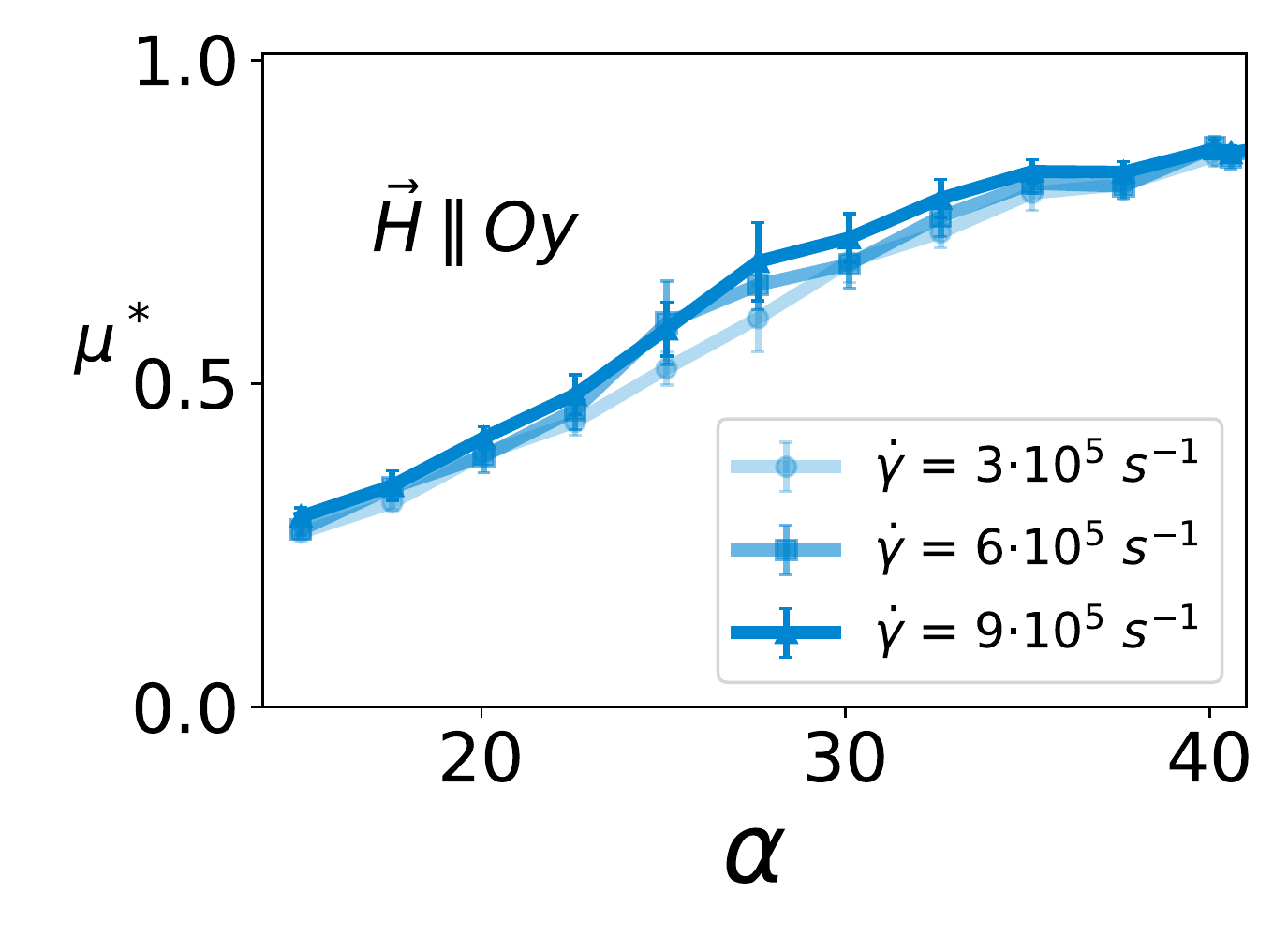}
            }
    \hfill
    \subfigure[]{\label{fig:MOz}
\includegraphics[ width=0.48\linewidth]{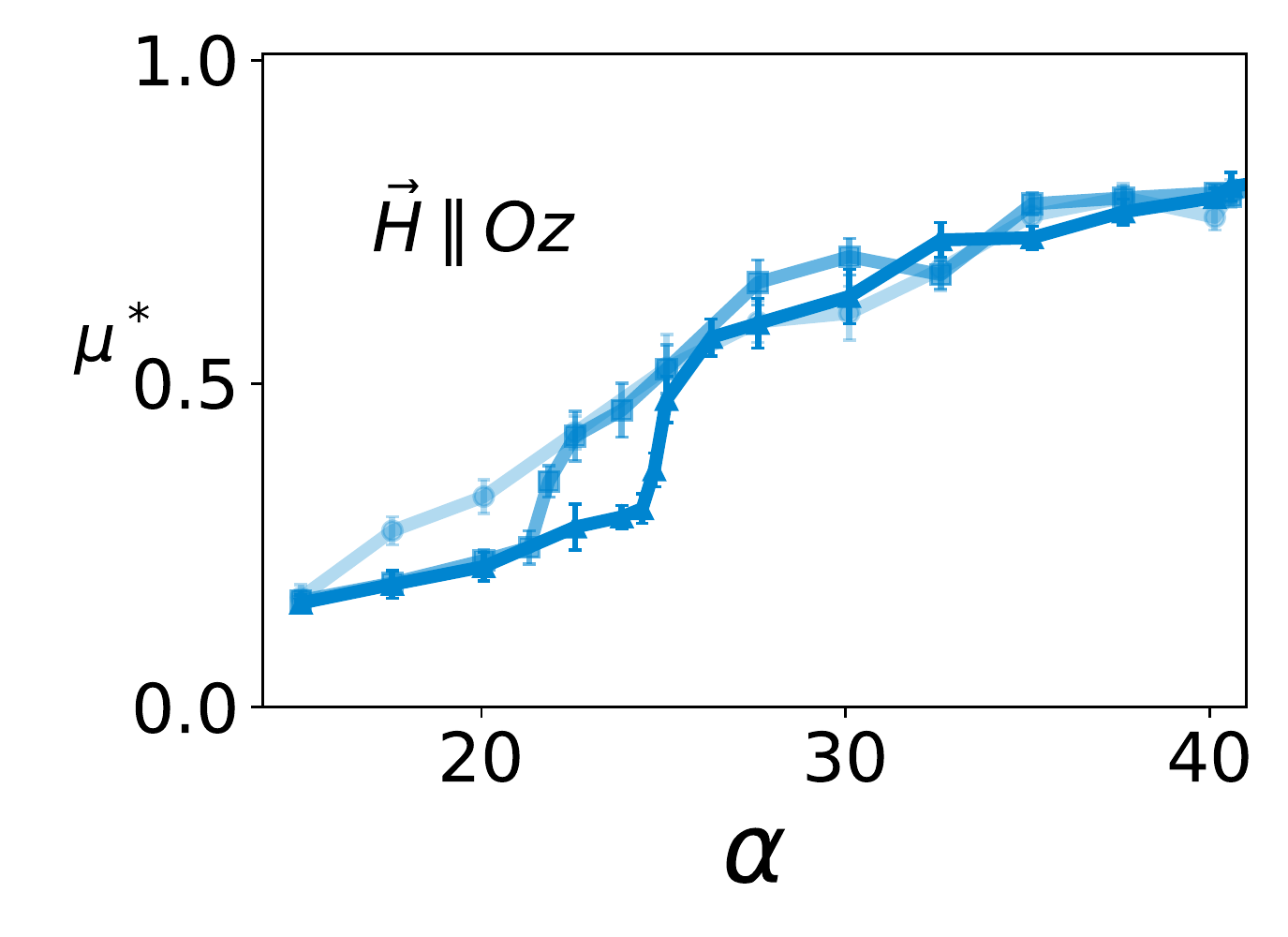}}
  \hfill
    \caption{Magnetisation $\mu^{\ast}$ for X-like cluster for the field aligned with Ox, Oy, Oz. The brightness of the curves corresponds to the value of the shear rate -- the darker the curve, the higher the value of $\dot{\gamma}$. Exact $\dot{\gamma}$ values and orientation of the applied field are provided in the insets.}\label{fig:x-magn}

\end{figure}

Looking at the orange and purple curves corresponding to the clusters made of  linear and ring-like SSMPs respectively, one can see that independently from the shear rate and the values of $\alpha$, magnetisation remains in a linear regime. This is not the case for the clusters made of X- and Y-like SSMPs: blue and gray curves have a characteristic s-shape, with two distinct regimes. Up to a certain value of $\alpha$, the regimes remain linear. For both X- and Y-like topologies, the linear regime persists longer for smaller shear rates with an overall magnetisation being higher for the clusters made of X-like SSMPs. For large values of $\dot{\gamma}$, the inflection point -- the value of $\alpha$ at which the linear regime changes -- is shifted towards stronger fields for the case clusters made of Y-like SSMPs. For the range of $\alpha$-values investigated here, only the clusters made of X-like SSMPs exhibit a crossover: the magnetisation in high shear reaches higher values than its counterpart for smaller $\dot{\gamma}$. This is clearly related to the behaviour of $Q$ discussed in previous sections.

In the middle plot for simplicity only the results for the highest and the lowest values of the shear rates are presented. Within four heat maps in Fig. \ref{fig:mag-vs-field-tot-ox}, instead, each of the three rows corresponds to a given value of the shear rate. The the darker the colour, the higher is the value of $\mu^{\ast}$. Horizontal axes show the values of the Langevin parameter, $\alpha$. One can see that independently from the topology of SSMPs forming the clusters, the darkest colour can be found in the lowest right corners, meaning the lowest shear rate and the highest field. An exception within the simulation error-bars can be found in the top left heat map that shows the magnetisation of the clusters formed by X-like SSMPs. In order to have a closer look at the magnetic properties of the latter clusters, in Fig. \ref{fig:x-magn}, we plot the dependence of $\mu^{\ast}$ {\it versus} $\alpha$ for all three orientations of an applied magnetic field and all investigated shear rates. It is particularly convenient to analyse those plots comparing them to those for the asphericity in Fig. \ref{fig:qh}.  In Fig. \ref{fig:MOx}, for the lowest shear rate, $\mu^{\ast}$ has no inflection, similarly to the lightest curve in  the left-most plot in Fig. \ref{fig:qh}. Thus, if the field  pointing along $Ox$ and $\dot{\gamma}$ is small, the cluster is steadily slowly elongating with growing $\alpha$ and its magnetisation follows the same trend. For a higher shear, both $\mu^{\ast}$ and $Q$ exhibit two regimes: up to $\alpha \sim 32$, both the elongation and the magnetisation grow slowly, then there is an inflection point within a short interval of a steep increase followed by the region where both parameters slowly reach the saturation.  For a high shear rate, the same s-shape growth of $\mu^{\ast}$ is also observed in the case of an applied field pointing along $Oz$, however, with an inflection point at a lower value of $\alpha \sim 25$, as shown in Fig. \ref{fig:MOz}. Here, for a small value of $\dot{\gamma}$ the shape of $\mu^{\ast}$ is very similar to its counterpart in Fig. \ref{fig:MOx}. Qualitatively different behaviour is found Fig. \ref{fig:MOy}, where $\vec{H}||Oy$. Here, independently from the shear rate, the magnetisation is found to grow first linearly and then a little slower. Moreover, the values of $\mu^{\ast}$ are higher if compared to the other two geometries. It is clearly caused by the fact that the cluster can afford a very strong in-field elongation and alignment if  $\vec{H}||Oy$, as seen in Fig. \ref{fig:q-orient}.

As long as the correlation between shape and magnetisation in a steady state is very pronounced, it is reasonable to expect the same interweave to manifest itself also in the time-evolution that is addressed in the next section.

\subsection{Magnetisation Dynamics}

\begin{figure*}[h!]
 \centering
\subfigure[]{\label{fig:MOx15}\includegraphics[width=0.26\textwidth]{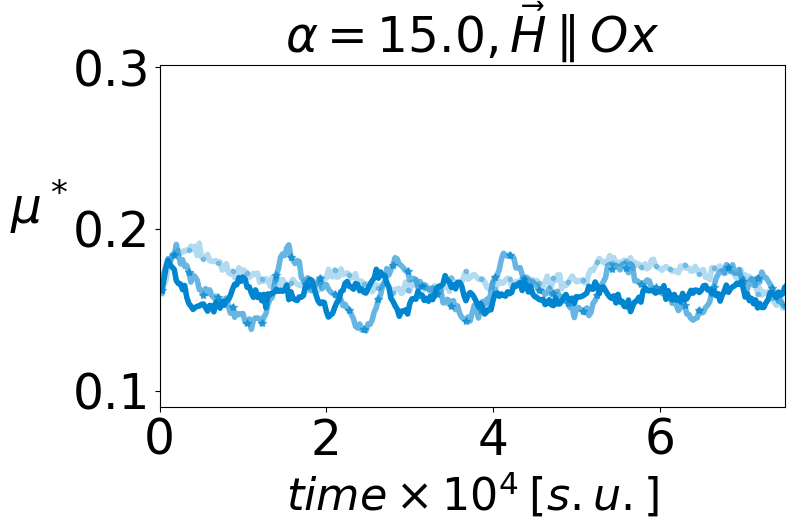}}
    \hfill
 \subfigure[]{ \label{fig:MOy15}\includegraphics[width=0.26\textwidth]{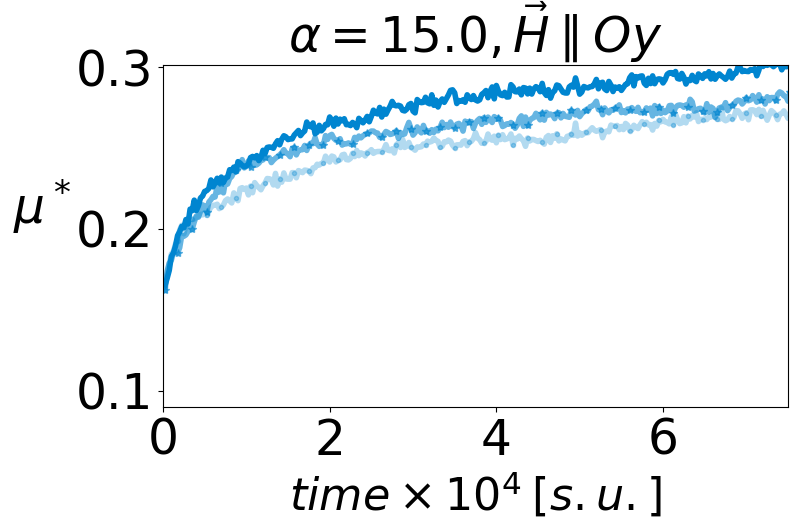}}
    \hfill
 \subfigure[]{\label{fig:MOz15}\includegraphics[width=0.26\textwidth]{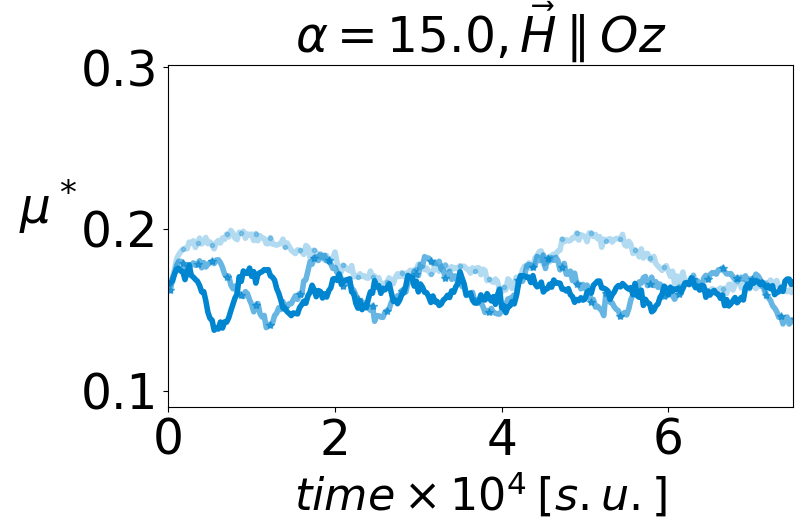}}
    \hfill
\subfigure[]{\label{fig:MOx30}\includegraphics[width=0.26\textwidth]{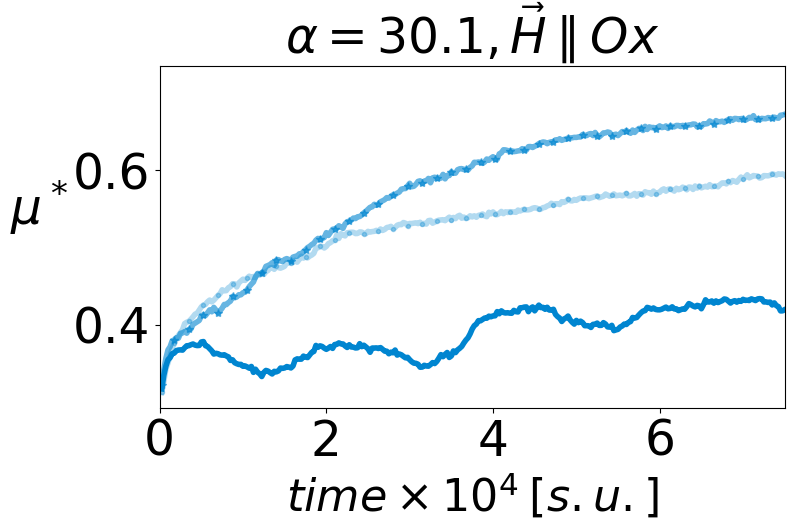}}
    \hfill
 \subfigure[]{\label{fig:MOy30}\includegraphics[width=0.26\textwidth]{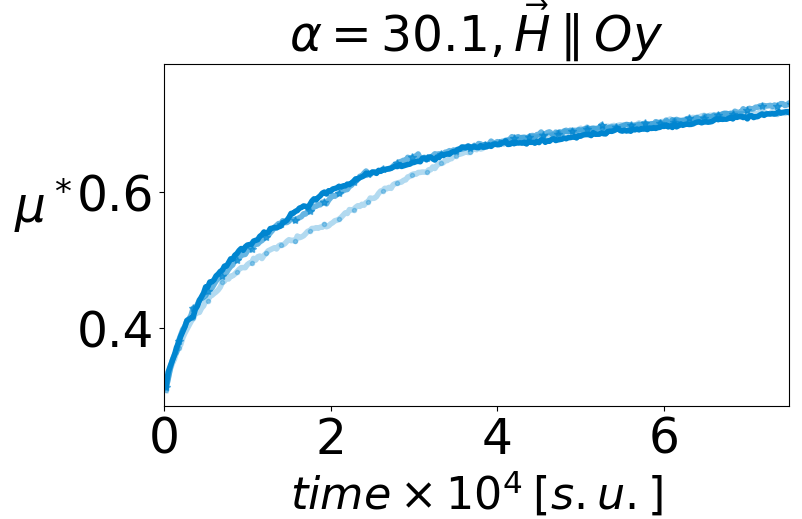}}
    \hfill
 \subfigure[]{\label{fig:MOz30}\includegraphics[width=0.26\textwidth]{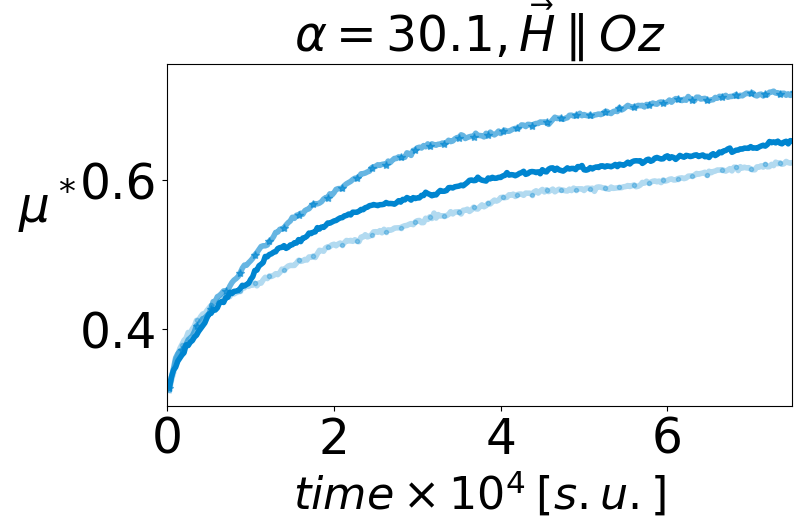}}
    \hfill
\subfigure[]{\label{fig:MOx42}\includegraphics[width=0.26\textwidth]{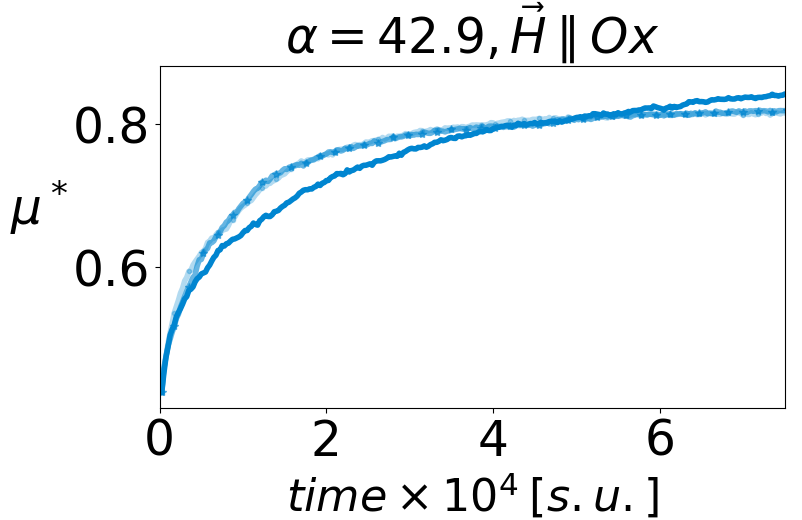}}
    \hfill
 \subfigure[]{\label{fig:MOy42}\includegraphics[width=0.26\textwidth]{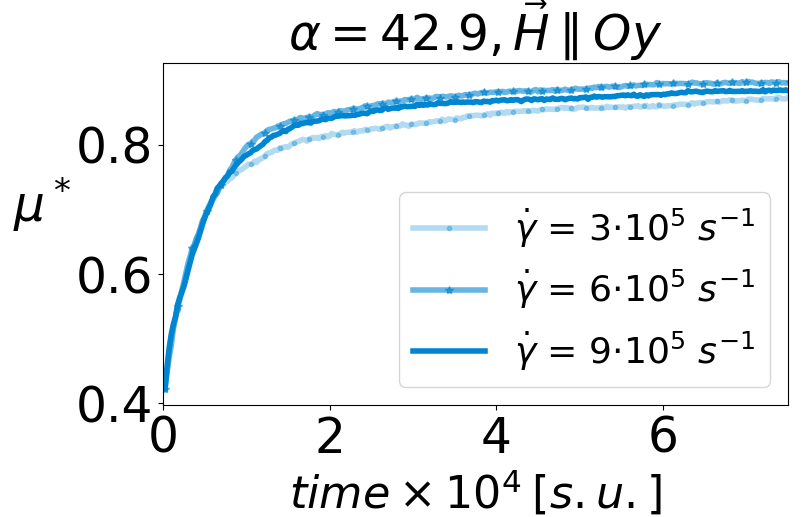}}
    \hfill
 \subfigure[]{\label{fig:MOz42}\includegraphics[width=0.26\textwidth]{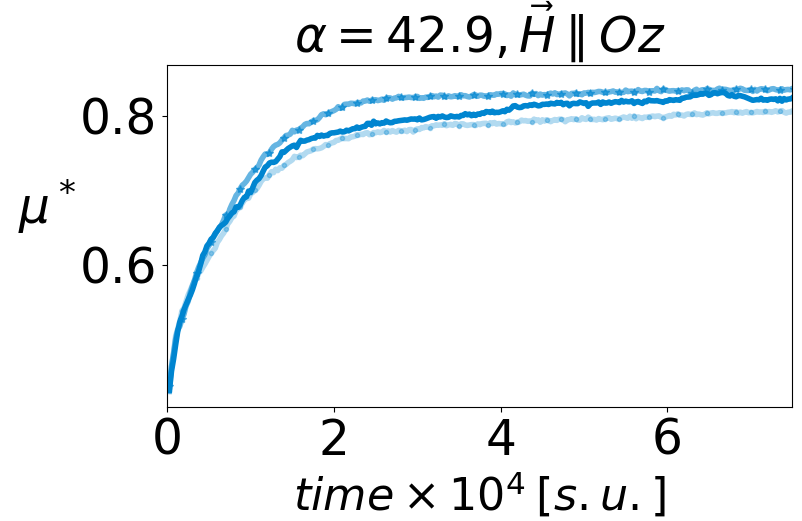}}
    \hfill
         \caption{Dynamics of the magnetisation $\mu^{\ast}$ for a X topology of the SSMPs. Columns correspond to a given orientation of the applied field. The brightness of the curves corresponds to the value of the shear rate -- the darker the curve, the higher the value of $\dot{\gamma}$. Exact $\dot{\gamma}$ values are provided in the inset of figure (h). Field is fixed within each row. Values of $\alpha$ are provided above the plots.}\label{fig:mag-dyn-x}

\end{figure*}

Similarly to the previous section, here we exclusively focus on the magnetisation dynamics of the clusters formed by X-like SMPs. In Fig. \ref{fig:mag-dyn-x} we present the dependence of cluster relative magnetisation, $\mu^{\ast}$, on the simulation time. Within each row the value of $\alpha$  is fixed and it grows from the top row to the bottom one. In each column the orientation of an applied field is preserved. For the upper row, both for  $\vec{H}||Ox$ and $\vec{H}||Oz$, see Figs. \ref{fig:MOx15} and \ref{fig:MOz15}, cluster magnetisation oscillates with similar frequencies that are growing with the shear rate.  In contrast, if an applied field is pointing perpendicular to both the flow velocity and the flow gradient,  $\vec{H}||Oy$, no oscillations of $\mu^{\ast}$ are observed as illustrated in Fig. \ref{fig:MOy15}. It is related to the stabilisation of the cluster shape with a pronounced elongation in the field direction. The middle row contains the results for $\alpha = 30.1$, see, Figs. \ref{fig:MOx30}--\ref{fig:MOz30}, the value close enough to the one presented in Figs. \ref{fig:x-like-qt-ox}--\ref{fig:x-like-qt-oz}. In fact, the only oscillations of $Q$ can be observed in Fig. \ref{fig:x-like-qt-ox} for $\dot{\gamma} = 9 \times 10^5 \ s^{-1}$, and the same behaviour is found for $\mu^{\ast}$, but with the frequency approximately twice lower, see Fig. \ref{fig:MOx30}. For the rest of the cases, Figs. \ref{fig:MOy30}--\ref{fig:MOz42}, nearly logarithmic growth of $\mu^{\ast}$ with time is observed. As expected from Fig. \ref{fig:x-magn}, the highest values of magnetisation are observed if the field is applied along $y$-axis. In the latter case only a weak dependence on the shear rate is found, albeit for the weakest field (Fig. \ref{fig:MOy15}), where the flow assists the elongation and leads to higher values of $\mu^{\ast}$.  

Summarising the discussion of magnetisation one can conclude that the clusters with the highest magnetic response are those that are made of SSMPs with junction, such as Y- ox X-like topologies. Moreover, the magnetisation of such clusters exhibits a high alignment with an applied magnetic field and as such is not subjected to the shear flow induced oscillations. Such a strong coupling to the field stabilises also an elongated shape of the cluster and prevents the oscillations, particularly, if compared to clusters made of linear or ring-like SSMPs. In other words, dipolar interactions combined with a central attraction lead to a very magnetically and structurally stable clusters that are not likely to be perturbed by the applied fields, while the presence of topological frustration in the building blocks makes the clusters more susceptible to magnetisation/elongation that in this case is also stabilised hydro-dynamically. 

\subsection{Comparison to other Types of Soft and Hard Colloids}
Although to the best of our knowledge there  no studies of soft magnetic polymer-based colloids with physical bonds are available in literature, it is important to underling both similarities and differences that these systems have with droplets or initially anisometric nano-sized composites subjected to a shear flow.

In the study of Cunha et al \cite{cunha18a} the authors investigated a ferrofluid droplet with a given surface tension, applied field oriented parallel and perpendicular to the flow velocity in a 2D channel subjected to a symmetric shear flow. Growing capillary magnetic number that they define as a product of the droplet magnetic permittivity and squared applied magnetic field and divided by the surface tension. Even though it is not possible to directly compare the deformation of the droplet to that of a cluster, as the capillary number cannot be directly defined for our systems, one can draw some qualitative conclusion: for the same relative change of the magnetic field value, the droplet elongates significantly more than a cluster, particularly, if the letter is made by linear- or ring-like SSMPs. One can attribute this effect firstly, to a stronger effective forces holding the cluster together and secondly, to the fact that while the droplet is liquid inside, the SSMP clusters are rather elastic due to precrosslinking. The extension of Cunha's investigations resulted in an exhaustive modelling of a 3D ferrofluid droplet in a symmetric flow with the mutual orientations of the flow and the field analogous to that we study here \cite{abicalil21a}. One can compare their field orientation along $z$ axis  to ours along $y$, and {\it vice versa}, while the flow velocity in both cases is coaligned with $Ox$. It is useful to compare the orientation of the cluster principle axis and cluster magnetisation for different geometries and SSMP topologies to the orientation of the droplet and its magnetisation, presented in Figs. 2 and 7 in Ref. \cite{abicalil21a}. Qualitatively the results a comparable, in particular the best alignment of the magnetisation along the field and the highest elongation of the droplet is observed if the field is oriented perpendicular both to the flow velocity and the flow velocity gradient. However, the dependence on the shear rates for a droplet show no crossover that we found for clusters made of X-like SSMPs in  Fig. \ref{fig:x-like-qt-oy}.  In general, similarly to the first case discussed in this section, the droplet is much more deformable than a cluster and as such its integrity cannot be stabilised by the flow.

If the colloids subjected to a combination of magnetic and hydrodynamic torques are hard, but initially elongated, the orientational distributions of the latter \cite{klop2016} seem to be rather different from the ones observed above and from those of deformable liquid droplets. 

In a study by Zablotsky \cite{ZABLOTSKY2019462} and previous works cited there in, the relationship between the viscosity of magnetic colloidal suspensions and an applied external magnetic field under shearing was investigated. The results revealed a significant increase in flow resistance due to the extension of internal structure induced by the magnetic field. In consequence of which, the overall viscosity of the suspension approximately doubled. Based on these findings, it is reasonable to anticipate a substantial increase in viscosity for a suspension of clusters made of X- and Y-like SSMPs when an external magnetic field is applied along the shear stress or flow velocity gradient.

\section{Conclusion}

In this study, we employed coarse grained Molecular Dynamics computer simulations in order to investigate cargo, transport and magnetic properties of the
clusters formed by chains, rings, X and Y shaped supramolecular magnetic polymers with magnetic single-domain sticky magnetic monomers (addressed to as SSMPs). To do so, we subject those cluster to a shear flow of varied intensity and an applied uniform magnetic field which direction can assume three different mutual orientations with the flow velocity.  Hydrodynamic interactions are modelled by coupling  Lattice-Boltzmann method to the molecular dynamics. 

We find that the clusters exhibit much higher stability than liquid droplets due to their complex internal interactions.  As such, those physically connected polymer-based clusters have overall promising cargo properties and  do not disintegrate in the flow up to rather high
values of an applied magnetic fields. The highest deformations, albeit lower than those observed for liquid droplets, are found for clusters formed by X- and Y-like SSMPs. Dynamics of a cluster shape in the flow has three distinct regimes: steady oscillation; oscillations with growing mean value; and oscillation free case. As long as  the clusters that exhibit the least oscillations are more
reliable for transport, as the periodic deformations might cause preliminary
and undesired release of the cargo, the clusters made of X- and Y-like SSMPs seem to be optimal, as their shape exhibits the least oscillations in the range of parameters analysed in this work. The same clusters were found to show the highest magnetic response and to align stronger with an applied magnetic field. Even though the simulations are too computationally costly to estimate the life-time of transition states, it is particularly surprising how long it takes for all clusters to disintegrate.

As the next step we will investigate the part of inter-cluster interactions on the magneto-rheology of their suspensions.

\section*{Acknowledgments}
This research has been supported by the RSF Grant No.19-72-10033. S.S.K. was partially supported by Project SAM P 33748. Computer simulations were performed at the Vienna Scientific Cluster (VSC) and at the Ural Federal University Cluster.

\bibliography{clusters-gels.bib, gels2021.bib, multicore.bib} 

\bibliographystyle{rsc} 
\end{document}